\documentclass[12pt]{article}
\usepackage{amsmath}
\usepackage{amssymb,amsfonts}
\usepackage[all]{xy}
\usepackage{graphicx}
\usepackage[parfill]{parskip}
\usepackage[usenames]{color}

\numberwithin{equation}{section}
\numberwithin{table}{section}
\def\beq{\begin{eqnarray}}
\def\eeq{\end{eqnarray}}

\newcommand{\ba}{\begin{eqnarray*}} 
\newcommand{\ea}{\end{eqnarray*}} 
\newcommand{\ban}{\begin{eqnarray}} 
\newcommand{\ean}{\end{eqnarray}} 
\newcommand{\be}{\begin{equation}} 
\newcommand{\ee}{\end{equation}} 
\newcommand{\ben}{\begin{equation}} 
\newcommand{\een}{\end{equation}}

\def\ep1{\epsilon_1}
\def\eps2{\epsilon_2}
\def\H{H_{{h}_0}^{h}}

\newcommand{\IZ}{\mathbb{Z}}
\newcommand{\IC}{\mathbb{C}}
\newcommand{\IP}{\mathbb{P}}
\newcommand{\IN}{\mathbb{N}}
\newcommand{\IR}{\mathbb{R}}

\newcommand{\nn}{\nonumber}

\newcommand{\cW}{{\cal W}}
\newcommand{\cN}{{\cal N}}

\newcommand{\cO}{{\cal O}}

\newcommand{\cF}{{\cal F}}

\newcommand{\cV}{{\cal V}}

\newcommand{\bk}{\mathbf k}

\newcommand{\Vs}{\cV}
\newcommand{\Vc}{V}
\newcommand{\Vt}{V}

\newcommand{\qs}{q_{s}}

\newcommand{\Tr}{{\rm Tr \,}}

\newcommand{\bkb}{{\bf \bar{k}}}
\newcommand{\bkp}{{\bf k'}}
\newcommand{\bkpb}{{\bf \bar{k}'}}

\begin{document}

\thispagestyle{empty}

\vskip 3cm
\noindent
{\LARGE \bf The toroidal block and the genus expansion}

\vskip .4cm
\begin{center}
\linethickness{.06cm}
\line(1,0){447}
\end{center}
\vskip .5cm
\noindent
{\large \bf Amir-Kian Kashani-Poor and Jan Troost}
 
\noindent

\vskip 0.15cm
{\em \hskip -.05cm Laboratoire de Physique Th\'eorique\footnote{Unit\'e Mixte du CNRS et
    de l'Ecole Normale Sup\'erieure associ\'ee \`a l'Universit\'e Pierre et
    Marie Curie 6, UMR
    8549.}}
    \vskip -0.22cm

{\em \hskip -.05cm Ecole Normale Sup\'erieure}
   \vskip -0.22cm

{\em \hskip -.05cm 24 rue Lhomond, 75005 Paris, France}

\vskip 1cm

\vskip0cm

\noindent {\sc Abstract:} 

We study the correspondence between four-dimensional supersymmetric gauge theories
and two-dimensional conformal field theories
in the case of ${\cal N}=2^*$ gauge
theory. We emphasize the genus expansion on the gauge theory
side, as obtained via geometric engineering from the topological
string. This point of view uncovers modular properties of the
one-point conformal block on a torus with complexified intermediate
momenta: in the large intermediate weight limit, it is a power series
whose coefficients are quasi-modular forms. 
The all-genus viewpoint that the
conformal field theory approach lends to the topological string yields insight into the
analytic structure of the topological string partition function
 in the field theory limit. 

\vskip 1cm

\pagebreak

{
\tableofcontents }

\newpage

\section{Introduction}
The two-dimensional / four-dimensional correspondence
\cite{Alday:2009aq}\cite{Wyllard:2009hg} relates observables and structural properties of
${\cal N}=2$ supersymmetric four-dimensional gauge theories to those
of two-dimensional conformal field theory \cite{Gaiotto:2009we}\cite{Drukker:2009tz}\cite{Alday:2009fs}\cite{Drukker:2009id}\cite{Bonelli:2009zp}. 
At the heart of the
correspondence lies the observation that the $\epsilon$-deformed
$\cN=2$ instanton partition functions \cite{Nekrasov:2002qd} map to
conformal blocks of conformal field theory. The gauge group and
field content of the gauge theory determine the worldsheet genus as well
as the number and weight of the insertions of the respective conformal
block. The partition sums on neither side of the correspondence are
known as analytic functions of their parameters. The correspondence
was established by comparing an expansion in the gauge coupling order
by order in instanton number on the gauge theory side to an expansion
in a complex structure parameter of the corresponding Riemann surface
with punctures on the conformal field theory side
\cite{Alday:2009aq,Poghossian:2009mk,Fateev:2009aw,Alba:2010qc}. However, the $\cN=2$
gauge theories can be geometrically engineered within string theory
\cite{Katz:1996fh} and this makes them amenable to worldsheet
techniques which give rise to the holomorphic anomaly equations
\cite{Bershadsky:1993ta,Bershadsky:1993cx}. The expressions one
obtains for the instanton partition function from this vantage point
are close in spirit to \cite{Seiberg:1994rs,Seiberg:1994aj}, namely,
they are exact in the gauge coupling, but obtained order by order in
an $\epsilon$ expansion. The underlying theme of this paper is to
contrast and exploit these two different expansions on the gauge
theory and the conformal field theory side. Though many aspects of
this paper will apply to a much larger class of theories, we will
concentrate here on the example of ${\cal N}=2^\ast$
$\epsilon$-deformed four-dimensional $SU(2)$ gauge theory (see also \cite{Huang:2012kn}). The
corresponding quantity on the conformal field theory side is the
one-point conformal block on the torus.

One of the hallmarks of both the Seiberg-Witten and the holomorphic
anomaly vantage point is the emphasis on modular properties, with the effective
coupling as modular parameter. It is by exploiting this symmetry that
exact expressions in the effective coupling can be obtained
\cite{Huang:2009md,Huang:2010kf,Huang:2011qx}. Modularity is of course
also a recurring theme on the conformal field theory side, but as a
property of $n$-point functions rather than of their constituent
conformal blocks. Imposing e.g. modular invariance of the torus
one-point function or crossing symmetry of the four-point function on
the sphere imposes constraints on sums over conformal
blocks. But if the correspondence is to hold beyond the weak coupling
regime, then each individual conformal block should have good modular
properties. This suggests that we should be able to construct
conformal blocks non-perturbatively with the appropriate complex
structure parameter serving as modular variable.
We indeed succeed at
extracting such results in the semi-classical (infinite central
charge) limit from null vector decoupling equations. That the
solutions of these equations exhibit quasi-modular behavior is
astounding. In studying the action of the modular group on the
equations and its boundary conditions, we are led to complexify the
exchanged momentum of the conformal block, corresponding to the vacuum
expectation value $a$ of the complex adjoint scalar field in the gauge
theory.  Enlarging the usual set of conformal blocks in this way is
required to allow for a natural action of the modular group.

Topological string theory, just as physical string theory, was
originally defined order by order in a genus expansion.
An exciting aspect of the two-dimensional / four-dimensional correspondence is that it provides a
non-perturbative definition of the topological string via the
corresponding conformal field theory -- albeit only on
certain geometries and in the field theory limit \cite{Dijkgraaf:2009pc,Cheng:2010yw,Mironov:2011dk}. In the following,
we will freely use the terminology of the topological string.
The action of the infinite dimensional chiral algebra on the conformal theory allows the derivation of recursion
relations in the complex structure parameter $q = e^{2 \pi i \tau}$ satisfied by conformal
blocks \cite{Zamolodchikov:1985ie,Zamo1986,Zamo1987} which are exact
in the $\epsilon$ parameters, and hence in the string coupling from a topological string perspective. This allows
us to derive all genus results for the topological string free energy
at any given order in $q$. Furthermore, these recursion relations
reveal a curious property regarding the structure of poles and zeros
of the topological string partition function $Z$. When we factorize the partition function as $Z =
Z_{pert} Z_{inst}$, with $Z_{pert}$ encompassing contributions not
involving base wrappings in the geometry underlying the engineering,
the recursion relations reveal a surprising infinite number of poles
of $Z_{inst}$ in the fiber class variable. These are exactly canceled
by zeros of $Z_{pert}$.

A characteristic of the Seiberg-Witten treatment of the problem, which
extends to the treatment via the holomorphic anomaly, is the presence
of two classes of variables: ultraviolet parameters such as the moduli space parameter $u$
or the bare coupling, and infrared parameters, such as the vacuum expectation value $a$ or
the effective coupling. The two-dimensional / four-dimensional correspondence only translates a
subset of these parameters into the conformal field theory context: the bare coupling
maps to a certain parameterization of the complex structure of the
punctured Riemann surface underlying the conformal block, whereas the parameter $a$
maps to the exchanged momentum. A natural question is to identify the
other parameters in the conformal field theory. In particular, the distinction between bare and effective coupling is noteworthy. The holomorphic anomaly equations yield results that are naturally modular in the effective coupling of the theory. In this paper, using conformal field theory methods, we will uncover modularity in $\cN=2^*$ amplitudes expressed in terms of the bare coupling (this is in the spirit of \cite{Minahan:1997if,Fateev:2009aw}). Generally, the choice of bare coupling constant is ambiguous. Indeed, in the other $SU(2)$ conformal theory, the theory with $N_f=4$
fundamental hyper multiplets, two natural definitions of the bare coupling are possible. We will discuss further in
a forthcoming paper how the choice with good modular properties is also distinguished from the conformal field theory point of view \cite{Marshakov:2009kj}.

The structure of this paper is as follows. We will review relevant
aspects of the torus one-point function and the topological string
partition function in section \ref{correspondence}, and introduce the
correspondence between two-dimensional and four-dimensional
observables. Section \ref{recursion} is dedicated to the recursion
relation satisfied by the torus one-point function, and lessons that
can be drawn for both the conformal field theory and the topological string by combining
it with modular results on the topological string side. In section
\ref{cftonepoint}, we analyze the constraint on the one-point function
on the torus arising from null vector decoupling. We solve the
resulting differential equation recursively in a semi-classical regime
and check the result against topological string theory results,
obtaining agreement. The modularity of the one-point function is
discussed in detail. We draw conclusions in section \ref{conclusions}.

\section{The one-point conformal block in the correspondence}
\label{correspondence}

In this section, we exhibit
the role of the one-point toroidal conformal block in conformal field
theory, discuss the corresponding quantity in topological string
theory, and then review how they are expected to match \cite{Alday:2009aq}.
We will recall the engineering of gauge theory within topological string
theory and freely use the language of the latter setup in the following.

Our formulas are based on the conformal algebra underlying any conformal
field theory. The only parameter which enters is the central charge $c$.
It will be useful however to also introduce the following
parameterizations  that have their origins in Liouville theory. For the
central charge, we set 
\be c = 1 + 6 Q^2 \,, \quad Q = b + \frac{1}{b}
\,.  
\ee
 The semi-classical Liouville limit $c \rightarrow \infty$ has the
incarnations $b \rightarrow 0$ or $b \rightarrow \infty$. In the
semi-classical limit, we connect to a classical Liouville theory
with action principle. To render an intuitive interpretation easier,
we sometimes parameterize conformal weights $h$ in terms of Liouville
momenta $\alpha$:

\be
h = \alpha (Q-\alpha) \,.  \ee

\subsection{The torus one-point function} \label{1_pt_review} We study
a two-dimensional conformal field theory of central charge $c$ on a
torus $\Sigma=T^2$ with modular parameter $\tau$. We will often write
$q=e^{2 \pi i \tau}$.
The toroidal one-point function of a Virasoro primary field $V_{h_m}$
with conformal dimension ${h}_m$ is also the traced cylinder
one-point function,
\be \langle \Vt_{{h}_m}
\rangle_\tau = \Tr \Vc_{h_m} q^{L_0 - \frac{c}{24}} \bar{q}^{\bar{L}_0
  - \frac{c}{24}} \,.  \ee 
The trace here is over a basis of the Hilbert space. In a conformal
field theory, this space is a direct sum of Verma modules associated
to the primaries $|h \rangle$ in the spectrum of the theory of weight
$h$. In principle, we need to indicate both the sum over left and
right conformal dimensions. For simplicity only, we suppose the
spectrum is diagonal.  Since the basis of primaries and descendents is
not orthogonal, we need to insert the overlaps of states when
expressing the trace of an operator in terms of its matrix elements. The
one-point function can thus be written as follows in terms of a basis of primaries
and descendents:
\be \langle \Vt_{{h}_m}
\rangle_\tau = \sum_h \sum_{{\bf k},{\bf \bar{k}},{\bf k'},{\bf
    \bar{k}'}} (M^h)_{(\bkp \bkpb)(\bk \bkb)}^{-1} \langle h | L_{\bf
  k} \bar{L}_{\bf \bar{k}} \Vt_{{h}_m} q^{L_0 - \frac{c}{24}}
\bar{q}^{\bar{L}_0 - \frac{c}{24}} L_{-\bf k'} \bar{L}_{-\bf \bar{k}'}
| h \rangle \,, \label{1_pt_tr} 
\ee 
where we have introduced the non-trivial overlaps of the descendent states,
 \be M^h_{(\bkp \bkpb)(\bk
  \bkb)} = \langle h| L_{\bf k} \bar{L}_{\bf \bar{k}} L_{-\bf k'}
\bar{L}_{-\bf \bar{k}'} | h \rangle \,.  
\ee 
We have used the
notation $L_{\bk} = L_{k_1} \ldots L_{k_m}$, with the sums running
over vectors of increasing dimension, and the components of the
vectors running over all positive integers.  By invoking the
commutator of the Virasoro generators with primary
fields,\footnote{\label{ops_on_cyl_and_plane}We distinguish notationally between operators on the cylinder, $\Vc_h$, and 
  on the plane, $\Vs_h$. With $\zeta = e^{-2 \pi i z}$, they are
  related by $\Vc_h(z) = (-2 \pi i \, \zeta)^h \Vs_h(\zeta)$.
 When we wish to
  refer to the operator without specifying the coordinate system, we
  use the notation $\Vc$ to avoid introducing a third symbol.}
\be
[L_n, \Vs_h(\zeta)] = \zeta^{n+1} \partial \Vs_h(\zeta) + h (n+1) \zeta^n \Vs_h(\zeta) \,,
\ee
 the matrix element in equation (\ref{1_pt_tr}) can be expressed in terms of
the three-point function
$\langle h| \Vs_{h_m}(\zeta) | h \rangle$ and its derivatives. The latter
can be evaluated explicitly, as the $\zeta$ dependence of the correlator
$\langle h| \Vs_{h_m}(\zeta) | h\rangle $ is fixed by conformal
invariance,\footnote{Note that the corresponding three-point function on the cylinder, $\langle h | \Vc_{h_m}(z) | h\rangle$, is independent of the insertion point $z$, as it should be.} 
$\langle h| \Vs_{h_m}(z) |
h\rangle \sim z^{-h_m}$. We hence obtain \be \sum_{{\bf k},{\bf
    \bar{k}},{\bf k'},{\bf \bar{k}'}} (M^h)_{(\bkp \bkpb)(\bk
  \bkb)}^{-1} \langle h | L_{\bf k} \bar{L}_{\bf \bar{k}} \Vt_{{h}_m}
q^{L_0 - \frac{c}{24}} \bar{q}^{\bar{L}_0 - \frac{c}{24}} L_{-\bf k'}
\bar{L}_{-\bf \bar{k}'} | h \rangle = (q \bar{q})^{{h}-\frac{c}{24}}
|{\cal F}_{{h}_m} ^{h} (q)|^2 \langle h | V_{h_m} | h \rangle \,.  \ee
The holomorphic quantity $\cF_{h_m}^h(q)$ is referred to as the
one-point conformal block on the torus. It encodes the contributions
of the descendants of a given primary to the one-point function. Note
that it is completely determined by the Virasoro algebra, hence
depends only on the quantities explicitly indicated and the central
charge of the conformal field theory. All of the dynamical information
of the conformal field theory is encoded in the three-point functions,
\begin{eqnarray}
C^{h}_{{h}_m,{h}} &=& \langle {h} | \Vc_{{h}_m} | {h} \rangle \,.
\end{eqnarray}
Expressed in terms of these quantities, the one-point function on the torus thus finally takes the form
\begin{eqnarray}
  \langle \Vt_{{h}_m} \rangle_\tau &=&
\sum_{{h}} C^{h}_{{{h}_m} ,{h}} (q \bar{q})^{{h}-\frac{c}{24}}
|{\cal F}_{{h}_m} ^{h} (q)|^2 \,.
\end{eqnarray}

The formalism we have described is very general.  In practice, the spectrum of
conformal field theories can differ significantly from theory to
theory.  Rational conformal field theories will have a finite
spectrum. Other theories can have discrete unitary spectra. Liouville
theory has a continuous unitary spectrum. There also exist non-unitary
conformal field theories with discrete or continuous spectra. 
All of these
theories will differ in their three-point functions, and in the set of
relevant conformal blocks.

\newpage

\subsection{The topological string theory} \label{subs_top}

The two-dimensional / four-dimensional correspondence relates conformal field theory to
gauge theory. The topological string enters our narrative as the
relevant gauge theories can be geometrically engineered within string
theory \cite{Katz:1996fh}. The tool we use to compute the gauge theory quantities is
the holomorphic anomaly equations \cite{Bershadsky:1993cx}, whose natural habitat is the
topological string. The holomorphic anomaly equations allow us to compute the gauge theory amplitudes in terms
of modular forms. These can be expanded to yield the instanton contributions to arbitrarily high order in the instanton number.

\subsubsection*{The topological string partition function }

The topological string partition function $Z$ is traditionally
assembled from the topological string amplitudes $F^g$ which are generating
functions for map counts from a Riemann surface of genus $g$ to the
target space $X$. The partition function weights the amplitudes with the string
coupling $g_s$,
\be Z= \exp \sum_{g=0}^\infty F^g g_s^{2g-2} \,.  \label{GW}
\ee 
In the limit of large K\"ahler parameters, the amplitudes have an expansion 
\be
F^g = \sum_{\mathbf k} d^g_{\mathbf k} \,Q_1^{k_1} \cdots Q_n^{k_n} \,,
\ee
 where the K\"ahler parameters $t_i = \int_{\Sigma_i} J$ are integrals of the
K\"ahler form $J$ on the space $X$, the parameters $Q_i$ are the
exponentials $Q_i = \exp( -t_i)$, and the surfaces $\Sigma_i$ furnish a basis of
homology two-cycles. The numbers $d^g_{\bf k}$ are rational, with
denominators encoding multi-wrapping contributions.
The sum is over curve classes labelled by the vector $\bk=(k_1,k_2,\dots,k_n)$.
 The Gopakumar-Vafa
form of the partition function re-expresses the partition function
$Z$ in terms of
  integer invariants $n^g_{\mathbf k}$,
\be Z = \exp \sum_g
\sum_{n=0}^\infty \sum_{\mathbf k} \frac{n^g_{\bf k} {\mathbf
    Q}^{n\mathbf k}}{n (\qs^{n/2}-\qs^{-n/2})^{2-2g} } \,, \label{GV} 
\ee
where the parameter $\qs= e^{g_s}$ is the exponential of the string
coupling and we have written ${\mathbf Q}^{\mathbf k} = \prod_i Q_i^{k_i}$. Given a curve class $\bk$, the invariants
  $n^g_\bk$ are zero for large enough genus $g$. Therefore, knowing a finite number of
  invariants $n^g_{\bk}$ yields the contribution of the curve class
  $\bk$ to the partition function to all genus. The two expansions (\ref{GW}) and
  (\ref{GV}) are related by invoking the formula
\be -\frac{1}{\sinh^2 x} = \frac{1}{x} + \sum_{k=1}^\infty \frac{2^{2k} B_{2k}}{(2k)!}
  x^{2k-1} \hspace{0.2cm}  \hspace{0.4cm}
\mbox{for} \quad
 x^2 < \pi^2 \,.  \ee

\subsubsection*{Geometric engineering}
The path from type
IIA string theory  to four-dimensional $\cN=2$ $SU(2)$ gauge theories
proceeds via six dimensions. Each two-cycle $\Sigma$ in the internal four-dimensional geometry gives rise to a perturbative
state corresponding to a massless photon. D2-branes and anti-D2 branes
wrapping the same cycle $\Sigma$ yield non-perturbative massive states
with mass $m$ proportional to the K\"ahler parameter of the cycle
divided by the string coupling, $m \sim \frac{t_\Sigma}{g_s}$. When
the cycle $\Sigma$ arises from the blowup of an $A_1$ singularity, the
perturbative and non-perturbative states are elements of the same
$SU(2)$ multiplet. The symmetry is broken by the non-vanishing size of
the cycle. In the blow-down limit $t_\Sigma \rightarrow 0$, the full
$SU(2)$ gauge symmetry is restored. To dimensionally reduce to four dimensions while
preserving $\cN=2$ supersymmetry, the compactification manifold must
be  Calabi-Yau. Fibering the $A_1$ singularity appropriately over a
$\IP^1$ gives rise to such manifolds. In the following, we refer to
the exceptional class resolving the $A_1$ singularity as the fiber
class $\Sigma_f$ and to the class of the $\IP^1$ the $A_1$ singularity
is fibered over as the base class $\Sigma_b$. The gauge coupling (squared) of
the four-dimensional theory is inversely proportional to the volume of
the compactification manifold, here the volume $t_{\Sigma_b}$ of
$\IP^1$ (in the compactification from six dimensions to four). The
weak coupling limit is therefore $t_{\Sigma_b} \rightarrow \infty$. Base wrapping number maps to instanton number in the gauge theory.
To retain worldsheet instanton corrections wrapping the base
(weighted by $e^{-t_{\Sigma_b}})$,  a gauge theory argument 
\cite{Katz:1996fh} shows that one needs to simultaneously scale $t_{\Sigma_f} \rightarrow
0$. To maintain W-bosons at finite mass, this requires scaling the
string coupling to zero as well. All in all, we can parameterize the field theory limit as follows:
\ba \label{parameters_ftl}
e^{-t_{\Sigma_b}} = \left(\frac{\beta \Lambda}{M_{string}} \right)^4 \,, 
\quad t_{\Sigma_f} = \frac{\beta a}{M_{string}} \,, 
\quad g_s = \beta \hat{g}_s \,, 
\ea
where we take $\beta \rightarrow0$. The scale $\Lambda$ enters via dimensional transmutation.
The power is determined by gauge theory considerations \cite{Katz:1996fh}. 
To add fundamental matter, one needs to blow up
points on the base $\IP^1$. Adjoint matter is obtained via partial compactification of the geometry \cite{Hollowood:2003cv}.

The topological string partition function on this class of geometries
can be computed via the topological vertex \cite{Aganagic:2003db}. For
the field theory limit to yield non-vanishing higher genus amplitudes,
the zeros from powers of the parameter $\beta$ in $g_s^{2g-2}$ must be
cancelled. This occurs via a resummation of contributions from fiber
wrappings. Using vertex techniques, this resummation can be performed
order by order in the base wrapping $k$. The calculation has been
performed in the case of $N_f=0$ for base wrapping number $k= 0,
\dots, 4$ in \cite{Iqbal:2003ix}, yielding the results 
\ban 
\sum_{n=1}^\infty \sum_{m} \frac{n^g_{(k,m)} {Q_b}^{n k} Q_f^{nm}}{n
  (\qs^{n/2}-\qs^{-n/2})^{2-2g} } &=&  \sum_{n=1}^\infty
\frac{P^g_{k}(Q^n_f)}{(1-Q^n_f)^{2g-2+4k}} \frac{ { Q_b}^{n k} }{n (\qs^{n/2}-\qs^{-n/2})^{2-2g} }
 \label{struc_resum} \\
&\longrightarrow_{\raisebox{0.4cm} \! \hspace{-0.6cm} \beta
  \rightarrow 0}&  \left(\frac{M_{string}}{a}\right)^{2g-2} P^g_{k}(1)\left(\frac{\Lambda }{a}\right)^{4k}  \hat{g}_s^{2g-2}   \label{pure}
\,.  
\ean 
Note that multi-wrapping contributions $n>1$ do not survive
the field theory limit. Explicit expressions for the polynomials
$P_k^g(x)$ can be found in \cite{Iqbal:2003ix}. 

It would be interesting to prove the pole structure in powers of the fiber parameter $Q_f$ in the 
resummation formula (\ref{struc_resum}) for all base wrapping numbers $k$ and for
more general gauge theories from the point of view of the vertex. In particular, the result for conformal theories, as determined via the holomorphic anomaly equations, differs from (\ref{pure}) in that $\Lambda/a$ is replaced by an $a$ independent factor $q_{inst}$,
\be
\left(\frac{\Lambda}{a} \right)^4 \rightarrow  q_{inst} \,.
\ee
The $a$ dependence of the amplitudes is hence independent of the instanton number $k$. This will be important for us in section \ref{top_less}.

\subsubsection*{The $\Omega$ deformation}
For topological strings on non-compact Calabi-Yau target spaces, it is
possible to introduce a second expansion parameter $s$ in the partition function,
\be Z=\exp
\sum_{n,g} F^{(n,g)}s^n g_s^{2g-2} \,. \label{refined} 
\ee
The conventional partition function is obtained by setting $s=0$. This
generalization goes under the name of $\Omega$ deformation. It was
first introduced in the field theory context in \cite{Nekrasov:2002qd} in a localization calculation of integrals over instanton moduli space. The two equivariant parameters $\epsilon_1$ and $\epsilon_2$ related to
spatial rotations in $\IR^4$ considered there are related to $g_s$ and $s$ via
\begin{eqnarray}
s&=& (\epsilon_1 + \epsilon_2)^2 \,,\nonumber \\
g_s^2 &=& \epsilon_1 \epsilon_2  \,.
\end{eqnarray}
The $\Omega$ deformation away from the field theory limit was studied
in \cite{Iqbal:2007ii} and related to a motivic count in
\cite{Dimofte:2009bv}. An interpretation of 
the free energies $F^{(n,g)}$ at $n\neq 0$ in terms
of map counts has not been put forward. The integrality properties of
the refined partition function are captured by relating it to a Schwinger
calculation along the lines of Gopakumar-Vafa. In the refined case,
variables $q_+$
and $q_-$ are introduced to keep track of both the left and the right
spin content of 
BPS states contributing to space-time graviphoton-curvature couplings. These are related to the parameters 
$\epsilon_i$ via the formulas $q_{\pm} = e^{-(\epsilon_1 \pm
\epsilon_2)}$.
In these variables, the refined partition sum takes the Gopakumar-Vafa form 
\begin{equation} 
Z=
\exp \displaystyle{\sum_{{2j_-,2j_+=0}\atop {n=1}}^\infty \sum_{\bk \in H_2(M,\mathbb{Z})} (-1)^{2(j_-+j_+)} \frac{N^{j_-j_+}_\bk } {n}   
\frac{\displaystyle{\sum_{m_-=-j_-}^{j_-}} q_-^{n m_-}}{2\sinh\left( \frac{n \epsilon_1}{2}\right)} \frac{\displaystyle{\sum_{m_+=-j_+}^{j_+}} q_+^{n m_+}} 
{2\sinh\left( \frac{n \epsilon_2}{2}\right)}e^{-n\, \bk \cdot t}} \,.
\end{equation} 
To relate the integers
$N^{j_-j_+}_\bk $ to the invariants $n^g_\bk$ discussed above, one
must introduce a particular basis in the space of spin representations,
\be
I^n=\left(2 [0]+\left[\frac{1}{2}\right]\right)^{\otimes n}=
\sum_{i}\left( \left(2n\atop n-i\right)- \left(2n \atop n-i-2\right)
\right)\left[\frac{i}{2}\right] \,,
\ee 
and sum over right moving spin setting $\epsilon_1=-\epsilon_2$,
\be 
\sum_{g=0}^\infty
n_{\bk}^g I^g=\sum_{j_+} N_{\bk}^{j_-j_+} (-1)^{2j_+}(2j_++
1)\left[j_-\right] \,.  
\ee 

The basis $I^n$ has the property
\be 
{\rm Tr}_{I^n} (-1)^{2 \sigma_3} e^{-2 \sigma_3 s}= (-1)^n \left(2 \sinh \frac{s}{2}\right)^{2n}\,,
\ee 
explaining the structure of formula (\ref{GV}).

\subsubsection*{Contributions with no fiber-wrapping}
For the two- / four-dimensional correspondence as reviewed below, it will be important to factorize the topological string partition function into a
 $q_{inst}$-independent perturbative factor $Z_{pert}$ and a $q_{inst}$-dependent factor $Z_{inst}$. The nomenclature is inspired by the interpretation
 in the field theory limit. Contributions to $Z_{pert}$ arise from maps that do not wrap the base $\Sigma_b$.

For the field theory limit of the topological string, an all genus
(and all orders in the deformation parameter $s$) expression is known for $Z_{pert}$. For the ${\cal N}=2^*$ theory,
 it is given by \cite{Nekrasov:2003rj,Hollowood:2003cv,Iqbal:2007ii}
\be \label{zpert}
\log Z_{pert} = \frac{1}{g_s^2} \log \frac{\Gamma_2(2a + m+
  \frac{\epsilon_1+\epsilon_2}{2}) \Gamma_2(-2a + m+
  \frac{\epsilon_1+\epsilon_2}{2}) }{\Gamma_2(2a) \Gamma_2(-2a)} \,.
\ee 
The function $\Gamma_2(x)$ is the Barnes' double Gamma function \cite{Barnes}, defined as the
exponential of the derivative of the Barnes' double zeta function $\zeta_2$,
\be
\Gamma_2(x| \epsilon_1, \epsilon_2) = \exp \frac{d}{ds} |_{s=0}
\zeta_{2} (s,x|\epsilon_1, \epsilon_2) \,, \ee 
\be
\zeta_2(s,x| \epsilon_1,\epsilon_2) = \sum_{m,n \ge 0} (m \epsilon_1 +
n \epsilon_2 +x)^{-s} = \frac{1}{\Gamma(s)} \int_0^\infty dt
\frac{t^{s-1} e^{-tx}}{(1- e^{-\epsilon_1 t}) (1- e^{-\epsilon_2 t})}
\,.  \ee 
The expression (\ref{zpert}) possesses an asymptotic expansion in the arguments
$g_s$ and
$s$ whose coefficients match  the results obtained via the
holomorphic anomaly equations. The expansion is Borel summable, but
not convergent. In section \ref{recursion} below, we will demonstrate that the two-dimensional / four-dimensional correspondence allows us to derive such all genus results to any order in the parameter $q_{inst}$.

\subsubsection*{The holomorphic anomaly equations}
The topological string results we will use in this paper were obtained
in \cite{Huang:2011qx} via application of the holomorphic anomaly
equations \cite{Bershadsky:1993cx}. These equations originate from
studying the worldsheet definition of the topological string, and
yield (together with appropriate boundary conditions) the topological
string partition function in the genus expansion (\ref{GW}). A
generalization of these equations was proposed in \cite{Krefl:2010fm,Huang:2010kf}
allowing to compute the refined amplitudes $F^{(n,g)}$ of equation
(\ref{refined}). Holomorphic anomaly equations can be derived that
 yield the partition function directly in the field theory limit.

\subsubsection*{The partition function of $\cN=2^*$ $SU(2)$ gauge theory}
In this paper, we concentrate on the example of $\cN=2^*$ $SU(2)$ gauge
theory. The holomorphic anomaly equations for this theory, and the
amplitudes $F^{(n,g)}$ they govern, depend on the gauge coupling
constant, the coordinate on moduli space $u$, and the adjoint mass $m$. As the theory is conformal in the massless limit, the identification of the
instanton expansion parameter $q_{inst}$ is difficult. For an in depth discussion, we
refer to \cite{Huang:2011qx}.

The amplitudes simplify dramatically in the massless case. The instanton expansion parameter $q_{inst}$ and the effective coupling can be identified in this case; they will simply be denoted as $q$ in the following. Also, the square of the scalar vacuum expectation
value $a$
becomes a good global coordinate on moduli space in this limit, proportional to the global coordinate $u$. Furthermore, as the expectation value $a$ becomes
the only massive parameter in the theory aside from the deformation
parameters $\epsilon_{1,2}$,
it can serve as a genus counting parameter. The amplitudes thus take
the form
\be F^{(n,g)} = \frac{1}{a^{2(g+n)-2}} \,
p_{(n,g)}\left(E_2(q),E_4(q),E_6(q) \right) \,.  \label{hol_massless} 
\ee 
The polynomials $p_{(n,g)}$ are homogeneous polynomials in the Eisenstein series of weight
$2(g+n)-2$. At order $g+n=2$, explicit expressions for the partition sums are \cite{Huang:2011qx}\footnote{With regard to the reference \cite{Huang:2011qx}, the normalization of the scalar vacuum expectation value is $a_{there}= 2a_{here}.$}
\begin{equation} 
 F^{(2,0)}=\frac{E_2}{768 \, a^2},\qquad  F^{(1,1)}=-\frac{E_2}{192 \, a^2},\qquad F^{(0,2)}=0,  
\end{equation} 
and at order $g+n=3$, one finds
\ba
F^{(3,0)}&=&-\frac{1}{368640 \, a^4}\left(5E_2^2+13E_4\right),\ \   
F^{(2,1)}=\frac{1}{184320 \, a^4}\left(25 E_2^2+29E_4\right)\,, \nonumber \\
F^{(1,2)}&=&-\frac{1}{15360 \, a^4}\left(5 E_2^2+E_4\right),\ \  F^{(0,3)}=0\, .
\ea
One can algorithmically generate the higher order terms.

\subsection{The correspondence}
The two-dimensional / four-dimensional correspondence in the case of $\cN=2^*$ $SU(2)$ theory identifies the torus one-point function with insertion of a primary field of dimension $h_m$ with the generalized instanton partition function of the gauge theory with adjoint matter of mass $m$. Parameters are matched as follows:
\begin{eqnarray}
b &=& \sqrt{\frac{\epsilon_2}{\epsilon_1}}  \,,
\nonumber \\
{h}_m &=& \frac{Q^2}{4} - \frac{m^2}{\epsilon_1 \epsilon_2}  \,,
\nonumber \\
{h} &=& \frac{Q^2}{4} - \frac{a^2}{\epsilon_1 \epsilon_2} \,.  \label{id_par}
\end{eqnarray}
This dictionary results in the identifications
\begin{eqnarray}
q^{h - \frac{c}{24}} {\cal F}_{{h}_m} ^{h} &=&   Z_{inst}   \,, \nonumber \\
C^{h}_{{h}_m,{h}} (\mbox{Liouville}) &=& Z_{pert} \, .   \label{id_z}
\end{eqnarray}
The first equality, involving conformal blocks, does not depend on the two-dimensional field theory
under consideration.  The second equality invokes the three-point function of Liouville
theory. One can think of the latter as the unitary theory with
continuous spectrum fixed by demanding null vector decoupling in its
correlation functions. As such, it is also governed by the
Virasoro algebra.

The second equality would seem to restrict the regime of validity of
the correspondence to the spectrum of Liouville theory. This would dictate strictly positive intermediate conformal dimensions
${h}$ and purely imaginary vacuum expectation value
$a \in i \, \mathbb{R}$. It will be interesting however to consider the first equality
for general values of complex intermediate conformal dimensions
${h}$. This is natural from the gauge theory point of view, where the vacuum expectation value $a$ is complex. Though this
takes us outside the realm of unitary conformal field theories, we
will see that it enables us to uncover interesting modular
structure in more general representations of the Virasoro
algebra. 

\section{Recursion, poles and modularity}
\label{recursion}
The technical heart of this section is the recursion relation
satisfied by the one-point conformal block on the torus. The conformal
block exhibits poles at degenerate weights; the recursion relation is
derived by determining the residues at these poles. In this section,
after reviewing this relation, we will study its consequences in light
of the two-dimensional / four-dimensional correspondence. Holomorphic
anomaly results will allow us to relate sums over residues that occur
in the recursion relation to modular expressions. In turn, we
demonstrate that the recursion relation order by order in $q$
provides all genus results for the topological string. Assuming
modularity, this provides an alternative to the holomorphic anomaly
equations for determining the topological string free energies.
 Finally, we point out a curious result regarding the zeros of $Z_{pert}$ and the pole structure of $Z_{inst}$
in the adjoint vacuum expectation value $a$.

\subsection{The recursion relation for the one-point conformal block}
In the semi-classical limit ${h},c,h_m \rightarrow \infty$ and $c/h$
and $h_m/h$ small (namely, the semi-classical approximation to the
integral over the momentum propagating in a given channel), the one-point toroidal conformal block simplifies to
\begin{eqnarray}
{\cal F}_{{h}_m}^{h} (q) & \rightarrow & \frac{q^{\frac{1}{24}}}{\eta(q)} \,.  \label{leading_one_point}
\end{eqnarray}
In the limit of large propagating conformal dimension $h$, the result is corrected by an infinite
power series in the modular parameter $q$ with coefficients decreasing
like negative powers of the exchanged conformal dimension
${h}$. The series is governed by a recursion relation derived in
\cite{Zamolodchikov:1985ie}\cite{Zamo1986}\cite{Zamo1987},
and most pedagogically in \cite{Hadasz:2009db}.

The correction term to the leading asymptotics can be captured by a function $H_{{h}_m}^{h}(q)$,
\begin{eqnarray}
{\cal F}_{{h}_m}^{h} (q)  &=& \frac{q^{ \frac{1}{24}}}{\eta(q)} H_{{h}_m}^{h} (q)  \,,
  \label{large_delta}
\end{eqnarray}
which satisfies a recursion relation.  We first define the
coefficients of its $q$-expansion: $H_{{h}_m}^{{h}}=\sum_{n=0}^\infty
H_{{h}_m}^{{h},n} q^n$. In this expansion, the poles in the propagating
conformal dimension ${h}$ are manifest. The expansion coefficients
satisfy the initialization condition and recursive formula
\begin{eqnarray}
H_{{h}_m}^{{h},0} &=& 1  \,,
\nonumber \\
H_{{h}_m}^{{h},n>0} &=&
\sum_{1 \le rs \le n} \frac{ A_{rs} P_{rs}(m) }{{h} - {h}_{rs}}
H_{{h}_m}^{{h}_{rs}+rs,n-rs}   \,, \label{rec_q}
\end{eqnarray}
where
\begin{eqnarray}
{h}_{rs} &=& \frac{Q^2}{4} - \frac{1}{4} (r b + s b^{-1})^2 \,,
\nonumber \\
A_{rs} &=& \frac{1}{2} {\prod_{\substack{   \raisebox{0.4cm}{} (p,q)=(1-r,1-s) \\    (p,q) \neq (0,0),(r,s) }    }^{(r,s) }\hspace{-0.9cm}{} }
\frac{1}{p b + q b^{-1}}   \,,
\nonumber 
\ean
\ban
 P_{rs}(m)  =  
\prod_{k=1}^{2r-1} \hspace{-0.9cm} \prod_{\substack{{l=1}  \\  \!\!\!\!  (k,l)= (1,1) \!\!\!\!\!\mod (2,2)}}^{2s-1} && \hspace{-1cm}
(\frac{m}{\sqrt{\epsilon_1 \epsilon_2}}+ \frac{k b + l b^{-1}}{2})
(\frac{m}{\sqrt{\epsilon_1 \epsilon_2}}+ \frac{k b - l b^{-1}}{2})
\nonumber \\
& & \hspace{-1cm}
(\frac{m}{\sqrt{\epsilon_1 \epsilon_2}}- \frac{k b + l b^{-1}}{2})
(\frac{m}{\sqrt{\epsilon_1 \epsilon_2}}- \frac{k b - l b^{-1}}{2}) \,.
\nonumber 
\end{eqnarray}
The recursion relation originates in the representation theory of the Virasoro algebra.
The solution to the recursion relation can be obtained order by order in the parameter
$q$ (and to high order)
using a symbolic manipulation program, and it can be successfully compared to the topological
string partition function $Z$. The instanton partition function
\cite{Nekrasov:2002qd} provides a solution to the
recursion relation in terms of sums over Young tableaux.

\subsection{The conformal field theory / topological string correspondence}
We now compare the conformal field theory and the topological string theory results on the partition function
$Z$ in the massless limit of ${\cal
  N}=2^\ast$ theory (i.e. ${\cal N}=4$ super Yang-Mills theory).
The conformal field theory point of view yields results that are exact
in the expectation value $a$ order by order in the modular parameter
$q$ (see equation (\ref{rec_q})), whereas the
holomorphic anomaly approach yields exact results in the parameter $q$ order by
order in the expectation value
$a^{-1}$ (see equation (\ref{hol_massless})).

To identify quantities on both sides of the correspondence, note that
an expansion of the recursion relation (\ref{rec_q}) in the large
expectation value limit is of the form
\be
H_{{h}_0}^{{h}} = 1 + \cO \left(\frac{1}{a^2} \right) \,.  
\ee
Given the dependence of the leading contribution
(\ref{leading_one_point}) to the one-point function on the expectation
value $a$ and the structure of the topological string partition
function (\ref{hol_massless}), we can make the identifications
\be \exp \sum_{n+g\le 1} \left[ F^{(n,g)}(a) \right]' s^n
g_s^{2g-2} = q^{{h}- \frac{c}{24}} \frac{q^{\frac{1}{24}}}{\eta(q)}
\ee 
and
\begin{equation}
H_{{h}_0}^{{h}} = \exp \sum_{n+g>1} \left[ F^{(n,g)}(a) \right]' s^n g_s^{2g-2} \,.   \label{n+g>1}
\end{equation}
The prime on the brackets $[\cdot]'$ indicates that terms constant in
the modular parameter $q$ have been dropped. 
In the topological string, these are 
the contributions from maps that do not wrap the base direction of the engineering geometry. 
They are captured by the Liouville three-point function (see equation (\ref{id_z})).

\subsubsection*{The $n+g \le 1$ contribution}

Using the formulas
\be
c = 1 + 6 \frac{s}{g_s^2}  \,, \quad {h} = \frac{1}{4} \frac{s}{g_s^2} - \frac{a^2}{g_s^2} \,,
\ee
we find
\be
q^{{h}- \frac{c}{24}} \frac{q^{\frac{1}{24}}}{\eta(q)} =  \exp (- \frac{a^2}{g_s^2} \log q - \log \eta) \,.
\ee
This allows us to determine the first terms in the topological string
partition function
\begin{eqnarray}
F^{(0,0)} &=& - a^2 \log q  \,, \\
F^{(1,0)} &=& 0  \,, \\
F^{(0,1)} &=& - \log \eta \,.
\end{eqnarray}
The part of the leading term which is physically significant in gauge theory
 (namely its $a$-dependence) matches
the expected behavior of the gauge theory prepotential of ${\cal N}=4$ super Yang-Mills
theory.\footnote{The $U(1)$ factor which enters the gauge theory side of the correspondence yields the semi-classical limit (\ref{leading_one_point}) of the one-point toroidal conformal block. For a general gauge theory, it modifies $F^{(n,g)}$ for $n+g \le 1$ by terms independent of the scalar vacuum expectation values.}

\subsection{A lesson for conformal field theory}

\subsubsection*{The $n+g >1$ contribution}
Exploiting the exact results for the partition sums $F^{(n,g)}$ obtained
via the holomorphic anomaly equations, we can use the identification (\ref{n+g>1}) to
obtain results to all orders in the modular parameter $q$ for the semi-classical
conformal block $H_{{h}_0}^{{h}}$. The
simplest such relation is obtained by comparing the order
$a^{-2}$ terms on both sides of equation (\ref{n+g>1}): 
\ba
- g_s^2 \sum_{1 \le rs \le n} q^{rs} A_{rs} P_{rs} (0) H_{{h}_{rs}+rs}  
&=& \left[ \frac{s^2}{g_s^2} F^{(2,0)} + s F^{(1,1)} + g_s^2 F^{(0,2)}\right]'\\
& =& \left[( \frac{s^2}{g_s^2} - 4 s ) \frac{E_2(q)}{384}\right]' \,.
\ea 
In the second line, we have used the explicit results from
\cite{Huang:2011qx}.

Note that we have found a surprising constraint on the residues
appearing in the conformal block recursion relation which is valid to
all orders in the modular parameter $q$. It moreover implies that
certain infinite sums over the residues have good modular
properties. We will return to the interpretation of this modularity
from the conformal field theory vantage point in section \ref{cftonepoint}. Comparing
higher order terms in $a^{-2}$ on both sides of the identification
(\ref{n+g>1})
yields an infinite set of such constraints. Order by order in $q$, the
validity of these constraints can be checked by invoking the recursion
relation (\ref{rec_q}). The constraints are more powerful than these perturbative
checks.

Finally, let us note that if we deform the ${\cal N}=4 $ theory to ${\cal N}=2^\ast$
through a mass deformation, we can continue the above exploitation of the correspondence,
order by order in $m/a$.

\subsection{Lessons for the topological string} \label{top_less}
\subsubsection*{Reconstructing the amplitudes from finitely many expansion coefficients in $q$}
Using the relation to conformal field theory and expanding the block
$\log \H$ perturbatively in the modular parameter $q$, we can derive all genus results for coefficients of the partition function $\log Z$ at any order in the parameter
$q$. E.g. to lowest order, we have 
\be \log \H = 1 - \frac{s (s - 4 g_s^2)}{8 g_s^2(
  4a^2-s)}q + \cO(q^2) \,, \ee which implies \be \left[\sum_{n+g>1}
  F^{(n,g)} g_s^{2g-2} s^n \right]_q = - \frac{s (s - 4 g_s^2)}{8
  g_s^2( 4a^2-s)} \,.  \label{examp_q} \ee 
The notation $[\cdot]_q$ indicates the coefficient of $q$ of the
quantity enclosed in the square brackets. By the structure of the
recursion relation (\ref{rec_q}) for $\H$, it is
evident that the coefficients of the monomial $q^n$ (where $n>0$) are rational functions
in the parameters $g_s$ and $s$, as explicitly exhibited here for the leading
term. This is in contrast to the $q$ independent contribution reviewed in section \ref{correspondence}
above.

Note that together with the knowledge that the topological string
amplitudes are quasi-modular, we can reconstruct the full amplitude at
a given order in the string coupling $g_s$ and the deformation
parameter $s$ by knowing a finite number of expansion coefficients in
the modular parameter $q$. As the dimension of the vector space of
quasi-modular forms of a given weight increases with weight, more
coefficients are necessary to reconstruct the partition function
$F^{(n,g)}$ at larger $n+g$.  At $n+g=2$ for instance, the polynomial
$p_{(n,g)}$ is of weight 2, and therefore proportional to $E_2$. The
coefficient of $\frac{s^2}{g_s^2}$ and of $s$ in equation (\ref{examp_q}) thus
completely determine the partition functions $F^{(2,0)}$ and $F^{(1,1)}$. To determine the
amplitudes at $n+g=3$, we require the coefficients of the forms $E_2^2$ and
$E_4$, and we must thus expand the block $\log \H$ to order $q^2$, etcetera.

\subsubsection*{Zeros and poles}
It is manifest that the recursion relation (\ref{rec_q}) for the
conformal blocks allows a resummation of the amplitudes
(\ref{hol_massless}). The increasing powers in $a^{-2}$ yield a
geometric series which is summed by expression (\ref{rec_q}). As can
be seen explicitly in the example (\ref{examp_q}), this gives rise to
poles in the $a$-plane at string coupling $g_s$ and deformation
parameter $s$ dependent positions. The occurrences of these poles is
unexpected from a physical point of view: singularities should arise
only where particles become massless or, from a geometric point
of view, when the target geometry is degenerating. In the case of the conformal field theory one-point function, the poles in the one-point conformal
blocks are cancelled by the zeroes in the three-point function. The same mechanism is at work here,
with the role of the three-point function played by the contributions to the partition function from curves not wrapping the base of the engineering geometry.

\section{The one-point function via null vector decoupling}
\label{cftonepoint}

In this section, we derive a differential equation that will allow us to determine the one-point conformal block on the torus, in the semi-classical limit $c \rightarrow \infty$, in an $\epsilon_1$ expansion. 

Following \cite{Belavin:1984vu}\cite{Eguchi:1986sb}\cite{Fateev:2009aw}, the strategy will be to insert an
 additional operator $V_h$ into the one-point function correlator
 $\langle \Vt_{{h}_m} \rangle_\tau$. By choosing $V_h$ to be
 degenerate, the resulting two-point function will be constrained by a
 null vector decoupling differential equation. Choosing the operator
$V_h$ to
 simultaneously be light in the semi-classical limit will allow us to
 extract the one-point function from this two-point
 function. While the one-point conformal block on the torus is a
 universal conformal field theory quantity depending only on the
 central charge of the theory, the extraction of the one-point from
 the two-point function in the semi-classical limit is most readily
 argued for in the context of Liouville theory.

The two-point function with one degenerate insertion corresponds to a
surface operator insertion in the gauge theory, or a brane insertion
in topological string theory \cite{Alday:2009fs,Teschner:2010je}.  In
\cite{Aganagic:2011mi}, following
\cite{Nekrasov:2009rc,Mironov:2009uv}, the open string topological
partition function is computed using matrix model techniques.  The
closed topological string partition function is then extracted from
the monodromies of the obtained result. Our approach to computing the
closed topological string partition function relies on the fact that
to leading order in $\epsilon_2$, the one-point (closed) and the
two-point (open) function coincide. The null vector decoupling
equation thus permits the direct computation of the one-point function
in the $\epsilon_2 \rightarrow 0$ limit.

\subsection{Heavy and light insertions in Liouville theory}
In this section, we follow the reasoning of \cite{Seiberg:1990eb}, as
reviewed and enhanced in \cite{Harlow:2011ny}, for the treatment of
Liouville theory in the semi-classical limit. The limit $c \rightarrow
\infty$ in any conformal field theory is referred to as
semi-classical. 
In the parameterization introduced at the beginning
of section \ref{correspondence}, this corresponds to the limit $b
\rightarrow 0$ (or $b \rightarrow \infty$). In this limit, primary
operators in Liouville theory can be expressed in terms of the
Liouville field $\phi$ via $V_\alpha = e^{2 \alpha \phi}$. Deviating
from the notation in the rest of this paper, we will label operators
in this subsection by their Liouville momentum $\alpha$ rather than their
conformal weight $h$.

The momenta of heavy operators in the semi-classical limit scale as
$\frac{1}{b}$ in the $b \rightarrow 0$ limit. Their insertion changes
the saddle point for the Liouville field. This effect can be
incorporated into a modified semi-classical action $S_L$. Light
operators have momentum scaling as $b$ and do not modify the semi-classical
saddle point. Their contribution to
the correlation function is multiplicative. In terms of the rescaled
field $\phi_{cl} = 2 b \phi$ which is finite in the $b \rightarrow 0$
limit, one thus obtains \cite{Harlow:2011ny} 
\be \langle
\Vs_{\frac{\eta_1}{b}}(\zeta_1,\bar{\zeta}_1) \cdots
\Vs_{\frac{\eta_n}{b}}(\zeta_n,\bar{\zeta}_n) \Vs_{b \sigma_1}(\xi_1,\bar{\xi}_1)
\cdots \Vs_{b \sigma_n}(\xi_n,\bar{\xi}_n) \rangle \approx e^{-\frac{1}{b^2}
  S_L[\phi_{cl}]} \prod_{i=1}^m e^{\sigma_i \phi_{cl}(\xi_i, \bar{\xi}_i)}
\,.  \ee 
Recall that we have parameterized $b =
\sqrt{\frac{\epsilon_2}{\epsilon_1}}$, and we are interested in the
one-point function $\langle \Vs_{\alpha_m} \rangle$ of momentum 
\be \alpha_m =\frac{Q}{2} - \frac{m}{\sqrt{\epsilon_1 \epsilon_2}} \,.  
\ee
To maintain the dependence on the mass
$m$ in the $b \rightarrow 0$ limit, we
consider the limit $\epsilon_2 \rightarrow 0$ while keeping
$\epsilon_1$ fixed. The operator $\Vs_{\alpha_m}$ is heavy in
this limit.

Degenerate operators $\Vs_{\alpha_{k,l}}$, parameterized by two positive integers $k,l \in \IN_0$, have Liouville momentum
\be
\alpha_{k,l} = \frac{Q}{2} - \frac{1}{2} \left(k\, b + \frac{l}{b} \right)  \,.
\ee
These operators are light for $l=1$. In the following, we will consider the lowest lying non-trivial such operator at $(k,l)=(2,1)$.
The semi-classical behavior of the corresponding two-point function is given by
\be \label{sc_factor}
\langle \Vs_{-\frac{b}{2}}(\zeta,\bar{\zeta}) \Vs_{\alpha_m}(0) \rangle  \approx  
e^{-\frac{1}{2} \phi_{cl}(\zeta, \bar{\zeta})}\langle \Vs_{\alpha_m}(0) \rangle \,,
\ee
where $e^{-\frac{1}{2} \phi_{cl}}$ is constant whereas $\langle \Vs_{\alpha_m}(0) \rangle$ scales as $e^{-\frac{1}{b^2}} \sim 
e^{-\frac{\epsilon_1}{\epsilon_2}}$ in the $\epsilon_2 \rightarrow 0$ limit.

\subsection{Isolating the contribution from a given channel} \label{iso}
As reviewed in section \ref{1_pt_review} in the case of the one-point
function, correlation functions on the torus can be defined as a trace
over the spectrum of the conformal field theory. In the case of
correlators with degenerate insertions, each summand will satisfy the
null vector decoupling differential equation separately. 
To isolate
particular summands, we study their
monodromy behavior upon circling the $A$-cycle of the torus,
i.e. under $z \rightarrow z+1$, following \cite{Mathur:1988yx}. Upon imposing this
behavior as a boundary condition, the differential equation has a
unique solution, which is essentially the conformal block we wish to
compute.

To determine the monodromy under $z \rightarrow z+1$, consider the correlator
\be \label{to_plane}
\langle h | \Vc_{h_m}(x)  \Vc_{(2,1)} (z) | h \rangle \sim \xi^{h_m} \zeta^{h_{(2,1)}} \langle h | \Vs_{h_m}(\xi)  \Vs_{(2,1)} (\zeta) | h \rangle \,,
\ee
where Greek letters indicate coordinates on the plane, $\zeta = \exp[ -2 \pi i \,z]$ 
etcetera.\footnote{See footnote \ref{ops_on_cyl_and_plane} for our conventions regarding the labeling of operators.} We denote the momenta corresponding to the weights $h_m, h_{(2,1)}, h$ as $\alpha_m, \alpha_{2,1}$, and $\alpha$ respectively. Considering only holomorphic dependence, we have
\ban
\Vs_{(2,1)}(\zeta) \Vs_h(0) &=& C_{(2,1), h}^{h_+} \zeta^{h_+ - h_{(2,1)} - h} \left( \Vs_{h_+}(0) + \beta_{(2,1),h}^{h_+} \zeta \,(L_{-1} \Vs_{h_+})(0) + \ldots \right) + \label{deg_ope} \\
&& + C_{(2,1), h}^{h_-} \zeta^{h_- - h_{(2,1)} - h} \left( \Vs_{h_-}(0) + \beta_{(2,1),h}^{h_-} \zeta \,(L_{-1} \Vs_{h_-})(0) + \ldots \right) \,. \nn
\ean
We have here exploited the fact that fusion with degenerate vectors
only involves a finite number of primaries \cite{Belavin:1984vu}. For
fusion with the vector $\Vs_{(2,1)}(z)$, the momenta of the two
primaries appearing in the operator product expansion are $\alpha_{\pm} = \alpha \pm
\frac{b}{2}$.
Using the parameterization (\ref{id_par}), we have
\be 
h_\pm - h = - \frac{b^2}{4} \pm \frac{ab}{\sqrt{\epsilon_1 \epsilon_2}}  \,.
\ee
{From} this, we read off the monodromy under $z \rightarrow z+1$ resulting from the two terms on the right hand side of (\ref{deg_ope}) inserted into equation
(\ref{to_plane}) to be
\be
\zeta^{h_\pm - h} \rightarrow   e^{\pi i \frac{b^2}{2} \mp 2 \pi i \frac{ab}{\sqrt{\epsilon_1 \epsilon_2}}}  \,.  \label{monod}
\ee
Imposing this monodromy as a boundary condition on the differential equation will prove to be sufficient to isolate the sought after contribution to the torus one-point function.

\subsection{The null vector decoupling equation}
We next turn to the derivation of the null vector decoupling equation on the torus 
\cite{Eguchi:1986sb}\cite{DiFrancesco:1987ez}\cite{Marshakov:2010fx}.
\subsubsection*{The conformal Ward identity on a torus}
To determine the appropriate differential equation, we need to recall the conformal Ward identity on the torus \cite{Eguchi:1986sb}. It is determined by combining Ward identities
for local reparameterization, local Lorentz and Weyl invariance, and involves an intricate cancellation
of potentially non-holomorphic contributions.
For the insertion of an energy-momentum tensor
in the correlation function of a product of vertex operators $\Vt_i$ inserted at points $z_i$, $i=1,2,\dots,n$, it
can be written as \cite{Eguchi:1986sb}

\begin{eqnarray}\label{wardtorus}
\lefteqn{\langle T(z) \prod_{i=1}^n\Vt_i(z_i) \rangle - \langle T \rangle \langle \prod_{i=1}^n \Vt_i(z_i) \rangle =}\\
&=& \sum_{i=1}^n \Big( h_i ( \wp(z-z_i)+ 2 \eta_1)  +(\zeta(z-z_i)+2 \eta_1 z_i) \partial_{z_i}
\Big) \langle \prod_{i=1}^n \Vt_i(z_i) \rangle 
 + 2 \pi i \partial_\tau \langle \prod_{i=1}^n\Vt_i (z_i)\rangle \,.  \nonumber
\end{eqnarray}
For simplicity, we have assumed that the torus has
periods $(1,\tau)$.  We have written (\ref{wardtorus}) in terms of the Weierstrass $\wp$-function and its primitive, the Weierstrass $\zeta$-function,
\be \label{defzeta}
\wp(z) =- \zeta'(z)   \,, \quad \zeta (z) = \frac{\theta_1'(z)}{\theta_1(z)} + 2 \eta_1 z \,,
\ee
and have introduced 
\be
\eta_1 = - \frac{1}{6} (\theta_1'''/\theta_1')|_{z=0} \,.
\ee

\subsubsection*{Imposing null vector decoupling}
The primary field $\Vt_{(2,1)}$ has a null vector at level two that
decouples from the conformal field theory by assumption. This implies
differential equations on correlators of the null vector with 
other vertex operators:
\begin{eqnarray}
\langle ((L_{-2} \Vt_{(2,1)}) (w)+ \frac{1}{b^2} (L_{-1}^2 \Vt_{(2,1)})(w)) \prod_{i=1}^n \Vt_i(z_i) \rangle
&=& 0 \,.
\end{eqnarray} 
We use the conformal Ward identity to compute the relevant operator product 
expansions between the energy-momentum tensor and the primary, and find
\ban
\langle (L_{-1} \Vt_{(2,1)})(w) \prod_{i=1}^n \Vt_i \rangle &=& \oint_w dz \langle T(z) \Vt_{(2,1)}(w) \prod_{i=1}^n \Vt_i \rangle   \nn\\
&=& \partial_w \langle \Vt_{(2,1)}(w) \prod_{i=1}^n \Vt_i \rangle \,, \label{lm1ontorus}
\ean
since only the $\zeta$ function exhibits a first order pole.  We also have
\ban
\lefteqn{\langle (L_{-2} \Vt_{(2,1)})(w) \prod_{i=1}^n \Vt_i \rangle = \oint_w \frac{dz}{z-w} \langle T(z) \Vt_{(2,1)}(w) \prod_{i=1}^n  \Vt_i \rangle }  \nn\\
&=& \Big[ 2  h_{(2,1)} \eta_1 + 2 \eta_1 w \partial_w + \sum_{i=1}^n \left( h_i ( \wp(w-z_i)+ 2 \eta_1)+(\zeta(w-z_i)+2 \eta_1 z_i)\partial_{z_i} \right) + 2 \pi i \partial_\tau \Big] \nn\\ &&\langle \Vt_{(2,1)}(w) \prod_{i=1}^n \Vt_i\rangle + \langle T \rangle \langle \Vt_{(2,1)}(w) \prod_{i=1}^n \Vt_i \rangle \,. \label{lm2ontorus}
\ean 
We have here used the conformal Ward identity (\ref{wardtorus}) with $n+1$
insertions, exploited that the residue of the function $\zeta(z)$
at zero is one, and that by their Taylor expansion the terms
$\wp(z-w)$ and $\eta(z-w)$ do not contribute. The calculation can
alternatively be thought of as an application of the conformal
algebra on the torus. We can simplify the null vector equation further by using the relation between the partition function $Z$
and the vacuum expectation value of the energy-momentum tensor on the torus,
$2 \pi i \partial_\tau \log Z = \langle T \rangle$. We then find
\begin{eqnarray}
\lefteqn{\Big[ \frac{1}{b^2} \partial_z^2
+ 2 \eta_1 z \partial_z +  \sum_{i=1}^n(\zeta(z-z_i)+2 \eta_1 z_i) \partial_{z_i}
} \nonumber \\
 &&+ 2 \pi i \partial_\tau + 2 h_{(2,1)} \eta_1
+ \sum h_i ( \wp (z-z_i) + 2 \eta_1) \Big] Z \langle V_{(2,1)} \prod_{i=1}^n V_i \rangle = 0 \,. 
\end{eqnarray}
We now apply the null vector decoupling equation to the two-point function
involving the degenerate field $V_{(2,1)}$ and one other insertion $V_{{h}_m}$. Invoking
the translation invariance of our conformal field theory on the torus, we simplify
the differential equation to
\be
\Big[ \frac{1}{b^2} \partial_z^2 +(  2 \eta_1 z -\zeta(z)   )\partial_{z}
+ 2 \pi i \partial_\tau + 2 h_{(2,1)} \eta_1
+  {h}_{m} ( \wp (z) + 2 \eta_1) \Big] Z \langle \Vt_{(2,1)} (z) \Vt_{{h}_m} (0)
\rangle =  0 \,.  \label{decoupling}
\ee
Further simplification can be achieved through the ansatz \cite{Fateev:2009aw}\cite{Marshakov:2010fx}
\be Z
 \langle \Vt_{(2,1)}(z) \Vt_{{h}_m} (0) \rangle_\tau = \theta_1(z|\tau) ^{\frac{b^2}{2}} \eta(\tau)^{2({h}_m - b^2 -1)} \Psi(z|\tau) \,.  
\label{diffeqansatz}
\ee
Plugging this ansatz into  equation (\ref{decoupling}), we obtain the differential equation \cite{Marshakov:2010fx}
\be \label{The_de_in_b}
\left[-\frac{1}{b^2} \partial_z^2 - \big(\frac{1}{4b^2} -
  \frac{m^2}{\epsilon_1 \epsilon_2} \big) \wp(z) \right] \Psi(z|\tau) = 2 \pi
i \partial_\tau \Psi(z|\tau) \,.  
\ee 
This form of the equation has the advantage of involving only operators and
functions that behave simply under modular transformations.

Note that the factor $\theta_1(z|\tau) ^{\frac{b^2}{2}}$ in the solution ansatz (\ref{diffeqansatz}) 
has a monodromy that
coincides with that of the channel independent contribution to the conformal block monodromy as determined in (\ref{monod}).

\subsection{The semi-classical conformal block}
\label{semiclassicalconformalblock}
In this section, we will provide an exponential ansatz for the
two-point function, project it onto the channel ${h}$, and solve the
differential equation (\ref{The_de_in_b}) perturbatively in the
semi-classical limit.
\subsubsection*{The exponential ansatz}
Substituting $b = \sqrt{\epsilon_2/\epsilon_1}$ into the differential equation (\ref{decoupling}) for the rescaled one-point function $\Psi$, we obtain 
\be \label{The_de}
\left[- \partial_z^2 - \big(\frac{1}{4} - \frac{1}{\epsilon_1^2}
  m^2\big) \wp(z) \right] \Psi(z|\tau) = \frac{\epsilon_2}{\epsilon_1}
2 \pi i \partial_\tau \Psi(z|\tau) \,.  
\ee 
This differential equation was derived in the strict $\epsilon_1
\rightarrow 0$ limit and solved in an $m/a$ expansion in
\cite{Fateev:2009aw}.
 Its $\epsilon_1 \rightarrow 0$ limit is similar
in form to a renormalization group equation for instanton corrections
to the ${\cal N}=2^\ast$ gauge theory \cite{D'Hoker:1997ha}.

Recall that we wish to solve equation (\ref{The_de}) in the semi-classical limit
$\epsilon_2 \rightarrow 0$, as it is in this limit that we can
disentangle the contribution of the auxiliary light insertion $V_{(2,1)}$ from
that of the heavy insertion $V_{h_m}$. The
factorization (\ref{sc_factor}) of the two-point function in this
limit suggests the ansatz
\begin{eqnarray} \label{ansatz_psi}
\Psi(z|\tau) &=& \exp \left[ \frac{1}{\epsilon_1 \epsilon_2} 
{\cal F}(\tau) + \frac{1}{\epsilon_1} {\cal W}(z|\tau)  + O(\epsilon_2) \right] \,,
\end{eqnarray}
such that 
\ban
\label{ansatz_F} \langle V_{h_m} \rangle &\approx& \exp
\left[-\frac{1}{b^2} S_L[\phi_{cl}] \right] \,\approx\, \exp
\frac{1}{\epsilon_1 \epsilon_2} {\cal F}(\tau) \,, \\
\label{ansatz_W} \exp \left[ -\frac{1}{2} \phi_{cl}(z, \bar{z})
\right] &\approx& \theta_1(z|\tau) ^{\frac{b^2}{2}} \exp
\frac{1}{\epsilon_1} {\cal W}(z|\tau) \,. \nn
\ean
 The powers of
$\epsilon_2$ in the ansatz (\ref{ansatz_psi}) are determined by the
$b$ scaling behavior of (\ref{sc_factor}). The leading $\epsilon_1$
behavior of $\cW(z|\tau)$ is motivated by the monodromy behavior
(\ref{monod}). 
Finally, the same leading behavior of $\cF(\tau)$ 
follows by requiring this monodromy to be compatible with the
differential equation (\ref{The_de}), as we will see below.  Note that
the factor $\eta(\tau)^{2({h}_m - b^2 -1)} $ in (\ref{diffeqansatz}),
with $\epsilon_{1,2}$ scaling behavior $\exp \cO(1)$, cannot be
unambiguously assigned to either $\langle V_{h_m} \rangle $ or $\exp
\left[ -\frac{1}{2} \phi_{cl}(z, \bar{z}) \right] $.

Plugging the ansatz into the differential equation (\ref{The_de}), we find
\be
-\frac{1}{\epsilon_1} \cW''(z|\tau) - \frac{1}{\epsilon_1^2} \cW'(z|\tau)^2 +  \left( \frac{1}{\epsilon_1^2} m^2 -\frac{1}{4} \right) \wp(z)  =(2 \pi i)^2 \frac{1}{\epsilon_1^2} q \partial_q {\cal F} (\tau)+  \frac{\epsilon_2}{\epsilon_1^2} 2 \pi i \partial_\tau \cW(z|\tau)  \,.   \label{diff_equ_exp}
\ee

\subsubsection*{Boundary condition from monodromy}
To project onto a channel with exchanged momentum $a$, we now impose
the monodromy behavior determined in equation (\ref{monod}). 

Since we already factored out $\theta_1^{b^2/2}$ in our ansatz (\ref{diffeqansatz}), accounting for the monodromy $e^{ \pi i b^2/2}$, we need to impose
the monodromy $e^{\pm 2 \pi i a b / \sqrt{\epsilon_1 \epsilon_2}}$ on
$\Psi$ under
the shift
 $z \rightarrow z+1$.  This translates into
 \be 
\cW(z+1) - \cW(z) = \pm 2\pi i a \,,
\ee
or equivalently, 
\be \oint \cW'(z) dz= \pm 2\pi i a \,.  
\ee 

\subsubsection*{Recursive definition of $\cW_n$ and $\cF_n$}
In the semi-classical approximation, we can neglect the term proportional to
$\epsilon_2$ in equation (\ref{diff_equ_exp}), and obtain
\be -\frac{1}{\epsilon_1}
\cW''(z|\tau) - \frac{1}{\epsilon_1^2} \cW'(z|\tau)^2 + \left(
  \frac{1}{\epsilon_1^2} m^2 -\frac{1}{4} \right) \wp(z) =(2 \pi i)^2
\frac{1}{\epsilon_1^2} q \partial_q {\cal F} (\tau)
\,.  \label{eq_in_w} 
\ee 
This is an ordinary first order differential equation for the
derivative $\cW'(z)$ depending on an unknown function ${\cal
  F}(\tau)$. If we perform a formal expansion of ${\cal F}$ and ${\cal W}$ in the parameter $\epsilon_1$,
\be
 {\cal F}(\tau) =
\sum_{n=0}^\infty {\cal F}_n(\tau) \epsilon_1^n \,, 
\quad {\cal  W}(z|\tau) = \sum_{n=0}^\infty {\cal W}_n(z|\tau) \epsilon_1^n \,,   \label{pert_ansatz}
\ee 
we obtain a system of equations for the coefficients $\cF_n$ and $\cW_n$,
\beq
-{\cW'_0}^2 + m^2 \wp &=& (2 \pi i)^2 q \partial_q \cF_0 \,, \label{first_eq} \\
-\cW''_0 - 2 \cW'_0 \cW'_1&=& (2 \pi i)^2 q \partial_q \cF_1 \,, \\
- \cW''_1 - {\cW'_1}^2  -2 \cW'_0 \cW'_2 - \frac{1}{4} \wp(z) &=&  (2 \pi i)^2 q \partial_q \cF_2 \,,\label{third_eq}\\
- \cW''_{n} - \sum_{i=0}^{n+1} \cW'_i \cW'_{n+1-i} &=& (2 \pi 
i)^2 q \partial_q \cF_{n+1}  \quad \mbox{for} \quad {n \ge 2} \,.
\label{generic_eq}
\eeq 
The projection onto the ${h}$ channel now reads
\be
\oint \cW_0' = \pm 2 \pi i a \,, \quad \oint \cW_i' = 0  \quad \mbox{for} \quad i>0
\,.  \label{bc} \ee

\subsection{The structure of the perturbative solution}

Consider the structure of the equations (\ref{generic_eq}) at a given
order $n$. Each increase in the order $n$ introduces a function
$\cW_{n+1}'$ which has not occurred in previous equations, as well as
a new function $q \partial_q \cF_{n+1}$ on the right hand side of the
equations. All other functions are known from lower order
equations. The strategy is to eliminate $\cW_{n+1}'$ from the equation by
integrating along the $[0,1]$ cycle and invoking the
boundary condition (\ref{bc}). We can thus determine $q \partial_q
\cF_n$ recursively for any $n$. By the two-dimensional /
four-dimensional correspondence, these modular expressions should
integrate to equal the topological string free energy $F^{(n,0)}$. We
checked this equality to high degree in $n$.\footnote{$\cF_n$ is
  determined up to a $\tau$-independent integration constant.} Note
that 
the fact that the $q$-dependence of the integral $\int d\tau\,
q \partial_q \cF_n$ for $n>2$ is captured by a polynomial or power series in
Eisenstein series (in the massless or massive case respectively) is
non-trivial by itself. 

For clarity of exposition,  we will treat the massless case first, 
before turning to the 
case of arbitrary mass.

\subsubsection*{The massless case}
In the massless case, the function $\cW_0'$ does not depend on the position
$z$ of the insertion. Integrating both
sides of the equation (\ref{generic_eq}) along the $[0,1]$ cycle thus
eliminates the unknown function $\cW_{n+1}'$ from the equation by
(\ref{bc}), allowing us to express $q \partial_q \cF_{n+1}$ in terms
of known quantities. This result in turn allows us to solve for the
function $\cW_{n+1}'$.

The first equation (\ref{first_eq}) has a slightly different structure
compared to the rest, as only here the unknown, $\cW_0'$, appears quadratically. We begin by solving the three
equations (\ref{first_eq})-(\ref{third_eq}) in the massless case. For simplicity, we will consider the
positive sign on the right hand side of the boundary condition (\ref{bc}) for
$\cW_0$ in the following. The other sign can be easily accessed via the Weyl transformation
$a
\rightarrow -a$. In Liouville theory this transformation corresponds to reflection symmetry. We obtain 
\be q \partial_q \cF_0 = - a^2 \,, \quad
\cW_0' = 2 \pi i a \,, \ee
 \be q \partial_q \cF_1 = 0\,, \quad
\cW_1'=0 \,, \ee and \be (2 \pi i)^2 q \partial_q \cF_2 = -\frac{1}{4}
\oint \wp(z) \,, \quad \cW_2' = \frac{1}{4} \frac{\oint \wp(z) -
  \wp(z)}{4 \pi i a} \,.  \ee 
The result for $\cW_0'$ follows from the
boundary condition, while the differential equation guarantees that
$\cW_0'$ is $z$-independent.

The manipulations we have outlined above, 
as well as the results on monodromies of Weierstrass functions recorded in appendix
\ref{monowp},
 will yield the functions
$\cW_n'$ as polynomials in the Weierstrass function $\wp(z)$ and its
derivatives with coefficients in the ring of quasi-modular forms
generated by the Eisenstein series $E_2, E_4, E_6$. We introduce a
grading on this space of solutions as follows: we assign weight 2 to
$\wp(z)$ and weight $2+n$ to $\wp^{(n)}(z)$, consistent with assigning
weights $2,4,6$ to the Eisenstein series $E_2, E_4, E_6$. The integral
$\oint$ is to commute with the grading. With this assignment, our
claim is that 
\be \cW_{2n}' = \frac{p_{2n}^e (\wp)}{a^{2n-1}} \,,
\quad \cW_{2n+1}' = \frac{\wp' \,p_{2(n-1)}^o (\wp)}{a^{2n}}
\,.  \label{w_general} \ee
 The polynomials $p_n^e$, $p_n^o$ are
homogeneous of weight $n$, 
\be p^e_{2n}(\wp) \,, p^o_{2n}(\wp) \in
\IC[E_2,E_4,E_6][\wp] \,.  \ee
 The proof follows easily by induction,
upon invoking the equations
 \beq
\wp'^2 &=& 4 \wp^3-g_2 \wp -g_3 \,,\\
\wp'' &=& 6 \wp^2 - \frac{1}{2}g_2 \,.  \eeq
 We note that the
derivative $q \partial_q \cF_{2n+1}$ is equal to the monodromy of the
derivative of a periodic function, and therefore vanishes.
We also have that
\be 
(2 \pi i)^2 q \partial_q \cF_{2n} = -\frac{1}{a^{2n-2}}\left( \sum_{i=0}^{n} \oint p_{2i}^{e}(\wp) p_{2(n-i)}^{e}(\wp) +\sum_{i=1}^{n-1}\oint p_{2(i-1)}^{o}(\wp) p_{2(n-i-2)}^{o}(\wp) \wp'^2\right) \label{f_general}  \,,
\ee 
which can be computed explicitly, recursively, and has weight
$2n$. Note that
 the expansion coefficients $\cF_n$ only depend on $a^2$.

The results up to $n=4$ are
\be
\partial_\tau \cF_2 = - i \frac{\pi}{24} E_2 \,, \quad \cW'_2 = i \frac{\pi^2 E_2 + 3 \wp}{48 \pi a}  \,,  \nonumber
\ee
\be
\partial_\tau \cF_3 = 0 \,, \quad \cW'_3 = - \frac{\wp'}{64 \pi^2 a^2} \,,  \nonumber
\ee
\be
\partial_\tau \cF_4 = i
\frac{\pi (E_2^2-E_4)}{4608 a^2} \,, \quad \cW'_4 = -i \frac{(2 E_2^2 - 25 E_4) \pi^4 + 6 \pi^2 E_2 \wp +225 \wp^2}{9216 \pi^3 a^3 }  \,,
\ee
and the next two non-vanishing orders of $\partial_\tau \cF_n$ equal
\be
\partial_\tau \cF_6 = -i \frac{\pi(5 E_2^3+21 E_2 E_4 - 26 E_6)}{1105920 a^4} \,, \quad
\partial_\tau \cF_8 = i \frac{ \pi (35 E_2^4 +329 E_2^2 E_4 - 1402 E_4^2 + 1038 E_2 E_6)}{297271296 a^6} \,. \nn
\ee
As promised, these integrate to elements of the polynomial ring generated by the Eisenstein series,
\be
\cF_4 = \frac{E_2}{768 a^2} \,, \quad
\cF_6 = - \frac{5 E_2^2+13 E_4}{368640 a^4} \,, \quad
\cF_8 = \frac{175 E_2^4+1092 E_2 E_4 + 3323 E_6}{743178240 a^6} \,.
\ee
These results coincide, via the identification
\be
\cF_{2n} = F^{(n,0)}
\ee
 with those obtained from the holomorphic anomaly equations \cite{Huang:2011qx} and quoted at the end of subsection \ref{subs_top}. It is easy to compute the amplitudes to higher order.

\subsubsection*{The massive case}

For non-vanishing mass, we expand all 
expansion coefficients $\cW_n'$ and $q \partial_q \cF_{2n}$ in powers of
\be
v =  \left( \frac{m}{2 \pi a} \right)^2 \,.
\ee

Solving for $\cW_{n+1}'$ and integrating both sides of the resulting
equations along the $[0,1]$ cycle, we can again eliminate the unknown
function $\cW_{n+1}'$ by invoking equation (\ref{bc}). Given $\cW_0'$
to a certain order in the parameter $v$, expanding $1/\cW_0'$ in the ratio $v$ permits solving
order by order for the coefficients of $q \partial_q \cF_{2n}$ in a
$v$ expansion.  This result in turn allows us to solve for the
function $\cW_{n+1}$. Assigning $v$ the weight $-2$, the structural
results (\ref{w_general}) and (\ref{f_general}) of the massless case
apply here as well, with $p^e_{2n}(\wp)$ and $p^o_{2n}(\wp)$ now power
series in the ratio $v$, with coefficients that are polynomials in $\wp$, with
coefficients in turn in the ring of Eisenstein series, 
\be
p^e_{2n}(\wp) \,, p^o_{2n}(\wp) \in \IC[E_2,E_4,E_6][\wp][[v]] \,.
\ee 
The leading term in the power series in $v$ is the massless
result.

To compute the function $\cW_0'$ (see also \cite{Fateev:2009aw}), consider again equation
(\ref{first_eq}), now for non-zero mass $m$. Together with the boundary
conditions, it yields the relations 
\be \oint_{\alpha} \sqrt{ (2
  \pi)^2 q \partial_q {\cal F}_0 + m^2 \wp} = 2 \pi i a \,, \quad
\cW_0' = \sqrt{(2 \pi)^2 q \partial_q \cF_0 + m^2 \wp }
\,.  \label{first_massive} 
\ee 
We will solve these equations in a
second perturbation series, in parallel to the $\epsilon_1$ expansion, in the
parameter $v$. Introducing the variable \be {\cal G} = -\frac{1}{a^2}
q \partial_q {\cal F}_0 \,,
\label{derivativeF}
\ee
we can rewrite the monodromy condition as
\begin{eqnarray}
\oint_{\alpha} \sqrt{ {\cal G} - v \wp} &=&  1.
\end{eqnarray}
For small $m/a$, we can expand the square root, and solve the equation order by order in $v$.
We suppose that ${\cal G}$ also has a series expansion $\sum_{n=0}^\infty {\cal G}^n v^n$ 
in terms of the parameter $v$. Using the Taylor series of the square root and
${\cal G}^0=1$, we then find
\begin{eqnarray}
\sum_{m=1}^\infty \frac{(-1)^m (2m)!}{(1-2m)(m!)^2 4^m} 
\oint
\left[ ({ \cal G}^1- \wp)v + \sum_{n=2}^\infty {\cal G}^n v^n  \right]^m
&=& 0.
\end{eqnarray}
At order $p$ in the parameter 
$v$, we obtain a linear equation on the coefficient ${\cal G}^p$, which can be solved
for in terms of the monodromies of powers of the Weierstrass function $\wp$. The latter
can again be solved for recursively, as demonstrated in appendix
\ref{monowp}.  We thus find that the coefficients ${\cal G}^p$ are
polynomials in the Eisenstein series of total weight $2p$, in accord
with our claim that $\cW_0$ have overall weight $0$.

The solutions up to fourth order in $v$ are the following:
\begin{eqnarray}
{\cal G}^0 &=& 1.
\nonumber \\
{\cal G}^1 &=&\oint_\alpha \wp = \frac{\pi^2}{3} (-E_2).
\nonumber \\ 
{\cal G}^2 &=& \oint_\alpha \frac{1}{4} ({\cal G}^1 - \wp)^2 = \frac{\pi^4}{36} (- E_2^2+E_4).
\nonumber \\
{\cal G}^3 &=& \oint_\alpha -\frac{1}{8} ({\cal G}^1 - \wp) (({\cal G}^1)^2 - 4 {\cal G}^2 - 2 {\cal G}^1 \wp
 + \wp^2)
\nonumber \\
&=& \frac{\pi^6}{540} (-5 E_2^3+3 E_2 E_4 + 2 E_6)
\nonumber \\
{\cal G}^4 &=&\oint_\alpha \frac{1}{64} (5 ({\cal G}^1)^4 - 24 ({\cal G}^1)^2 {\cal G}^2 + 16 ({\cal G}^2)^2 
+ 32 {\cal G}^1 {\cal G}^3
 - 20 ({\cal G}^1)^3 \wp + 
\nonumber \\
& & 
  48 {\cal G}^1 {\cal G}^2 \wp - 32 {\cal G}^3 \wp + 30 ({\cal G}^1)^2 \wp^2 - 24 {\cal G}^2 \wp^2 - 20 {\cal G}^1 \wp^3 + 
  5 \wp^4)
\nonumber \\
 &=& \frac{\pi^8}{9072} (-35 E_2^4 + 7 E_2^2 E_4 + 10 E_4^2 + 18 E_2 E_6).
\end{eqnarray}
These results in turn allow us to compute $\cW_0'$ in a $v$ expansion,
\be \cW_0' = 2 \pi i a ( 1 - \frac{v}{6} ( \pi^2 E_2 + 3 \wp) -
\frac{v^2}{72} ( 2 E_2^2 \pi^4 - E_4 \pi^4 + 6 E_2 \pi^2 \wp + 9
\wp^2) + O(v^3)) \,.  \ee With $\cW_0'$ in hand, the above procedure
then readily yields the higher $\cW_n'$ and $\cF_n$ recursively. We
list here results for the first few orders in $\epsilon_1$ and $v$,
\begin{eqnarray}
\cW_1' &=& \frac{v}{4} \wp' + \frac{v^2}{12} ( 3 \wp \wp' + E_2 \pi^2 \wp') 
+ O(v^3) \,,
\nonumber \\
\partial_\tau {\cal F}_2 &=& - \pi i \frac{1}{24} E_2 - v \pi^3 i \frac{1}{144} (E_2^2-E_4)
+ O(v^2)\,,
\nonumber \\
\cW_2' &=& \frac{i}{48 a \pi}
( E_2 \pi^2 + 3 \wp)
+ \frac{i}{288 a \pi}
( 2 E_2^2 \pi^4 - 13 E_4 \pi^4 + 6 E_2 \pi^2 \wp +
117 \wp^2) v + O(v^2)\,,
\nonumber \\
\cW_3' &=& -\frac{1}{64 a^2 \pi^2}
\wp'
- 
\frac{1}{96 a^2 \pi^2}
(E_2 \pi^2 + 21 \wp) \wp' v+O(v^2)\,,
\nonumber \\
\partial_\tau {\cal F}_4 &=& 
\pi i \frac{1}{4608 a^2} (E_2^2-E_4)
+ \pi^3 i \frac{1}{23040 a^2} ( 5 E_2^3 + 13 E_2 E_4 - 18 E_6) 
v
+ O(v^2)\,,
\nonumber \\
\cW_4' &=& 
- \frac{i}{9216 a^3 \pi^3}
\Big(( 2 E_2^2 - 25 E_4) \pi^4 + 6 E_2 \pi^2 \wp+
225 \wp^2 \Big) 
\nonumber \\
& & 
- \frac{i}{276480 a^3 \pi^3}
\Big((60 E_2^3 + 657 E_2 E_4 - 4748 E_6) \pi^6 + 45 \wp
(4 E_2^2 - 705 E_4) \pi^4 
\nonumber \\
& & 6885 E_2 \wp^2 \pi^2 + 160245 \wp^3 \Big) v + O(v^2) \,.
\end{eqnarray}
Following \cite{Fateev:2009aw}, we can use formula (\ref{derivativeF}) to integrate for the leading expansion coefficient $\cF_0$.
We find
\begin{eqnarray}
{\cal F}_0 &=& - a^2 \log q + 2 m^2 \log \eta
+ \frac{1}{48} \frac{m^4}{a^2} E_2
+  \frac{1}{5760} \frac{m^6}{a^4} (5 E_2^2+E_4)
\nonumber \\
& & 
+ \frac{1}{2903040} \frac{m^8}{a^6} (175 E_2^3 + 84 E_2 E_4 + 11 E_6)
+ O((\frac{m}{a})^{10}) \,.
\end{eqnarray}
This coincides up to normalization with the prepotential determined in
\cite{Minahan:1997if} via Seiberg-Witten techniques. We can go beyond
this result by integrating the expansion coefficients $\cF_{2n}$,
obtaining expressions for the couplings $F^{(n,0)}$ in the dual gauge
theory:
\be
\cF_2 = - \frac{\log \eta}{2} - \frac{E_2}{96} \frac{m^2}{a^2} + O((\frac{m^2}{a^2})^2)  \,,
\ee
\be
\cF_4 = \frac{E_2}{768 a^2} + \frac{(5 E_2^2 + 9 E_4)}{7680 a^2}\frac{m^2}{a^2} + O((\frac{m^2}{a^2})^2)  \,,
\ee
\be
\cF_6 = - \frac{5 E_2^2 + 13 E_4}{368640 a^4} - \frac{(35 E_2^3 + 168 E_2 E_4 + 355 E_6)}{9289728 a^4} \frac{m^2}{a^2} + O((\frac{m^2}{a^2})^2) \,.
\ee
These solutions are perturbative in the parameter $m/a$, and non-perturbative in the modular parameter $q$. 
In the massive case, we generate an infinite
series of primitives of polynomials in the Eisenstein series $E_2$,
$E_4$, and $E_6$ for each expansion coefficient $\cF_n$, $n>2$. Note that the same amplitudes were obtained in \cite{Huang:2011qx} in closed form in terms of Eisenstein series in the effective coupling.

\subsubsection*{Changing the boundary condition} \label{changing_bc} 
In this subsection, we want to analyze the role of the choice of cycle in
the boundary conditions.  It is clear from the modularity of the
differential equation that modular transformations act on the space of
solutions of the differential equation (see e.g. \cite{Zhu}). It is
therefore natural to ask how the boundary condition satisfied by a
given solution changes under modular transformations.

To answer this question, let us introduce a basis of cycles
$(\alpha,\beta)$ of the torus, such that in the lattice
representation, the path $[0,1]$ is a representative of $\alpha$ and
$[0,\tau]$ a representative of $\beta$. We denote a solution of the
differential equation (\ref{eq_in_w}) by 
\be ( w_{\gamma,a},
f_{\gamma, a}) \,, \ee 
where 
\be \cW(z|\tau) = w_{\gamma,a}(z|\tau)
\,, \quad \cF(\tau) = f_{\gamma,a} (\tau) \,,
\ee
and we impose the boundary condition on the cycle $\gamma$
\be
\oint_\gamma \partial_z w_{\gamma,a} (z|\tau) \, dz = 2\pi i a(\tau)
\,.  \ee 
Note that by an extension of our above analysis to general
boundary conditions, the function $\partial_z w_{\gamma,a}$ has no
residue, and hence the integral with regard to a cycle is
well-defined.

Given a solution $(w_{\alpha,a},f_{\alpha,a})$ associated to the $\alpha$ cycle, define new functions:
\be
w(z|\tau) = w_{\alpha,a}\!\left( \frac{z}{c\tau+d} \Big| \frac{a \tau +b}{c\tau+d} \right) \,, \quad f(\tau) = f_{\alpha,a}\! \left( \frac{a \tau +b}{c\tau+d} \right)  \,. \label{new_from_old}
\ee
 In the following, we write
\be 
z' =  \frac{z}{c\tau+d} \,, \quad \tau' = \frac{a \tau +b}{c\tau+d}=g(\tau) \,.
\ee
By
\be
\partial_z w(z|\tau) = \frac{1}{c\tau+d} \partial_{z'}
w_{\alpha,a}(z'|\tau') \,, \quad \partial_\tau f(\tau) = \frac{1}{(c
  \tau + d)^2}\partial_{\tau'} f_{\alpha,a}(\tau') \ee and taking into
account the transformation properties of the Weierstrass function
$\wp$, we see that the pair $(w,f)$
is a solution to the differential equation (\ref{eq_in_w}) as well. To
determine the boundary condition, we compute
\be
\oint_{d \alpha + c  \beta} \!\!\!dz\, \partial_z w(z|\tau) = \int_0^{c\tau+d}\!\!\!dz\, \partial_z w(z|\tau) = \int_0^1 dz'\, \partial_{z'}
w_{\alpha,a}(z'|\tau') 
= 2\pi i a(\tau') \,.  
\nonumber 
\ee
Defining a cycle $\gamma = d \alpha + c \beta = g(\alpha)$, we can hence identify the solution $(w,f)$ as 
\be (w,f) = (w_{\gamma,a \circ \tau'}, f_{\gamma,a
\circ \tau'}) \,.  
\ee 
We conclude that the action of $SL(2,\mathbb{Z})$ on the space of solutions is given by 
\begin{eqnarray}
g \in SL(2,\IZ) \,: \quad  (w_{\alpha,a},f_{\alpha,a}) & \mapsto& (w_{g(\alpha),a \circ g},f_{g(\alpha),a \circ g}).
\end{eqnarray}
When the function $a(\tau)$ does not depend on $\tau$, it is always the same
constant  that appears on the right hand side of the boundary
conditions on the various cycles $\gamma$.

\subsubsection*{Colliding insertions}
Finally, we want to address a subtle point.
A glance at equation (\ref{eq_in_w}), keeping in mind the double pole of the
Weierstrass $\wp$ function at $z=0$, convinces us that the $z
\rightarrow 0$ limit and the $\epsilon_1 \rightarrow 0$ limit are not
independent. Our perturbation ansatz (\ref{pert_ansatz}) tacitly
assumed $z$ fixed away from $z=0$. Indeed, the perturbative solution
we derive diverges as $z \rightarrow 0$.\footnote{We would like to thank Don Zagier
for stressing this point.}
In this limit, as the degenerate operator approaches $V_{h_m}$, the operator product expansion of the two operators suggests the behavior
\be
\langle \Vt_{(2,1)}(z) \Vt_{{h}_m} (0) \rangle_\tau \underset{z \rightarrow 0}{\sim} z^{-h_m-h_{(2,1)}+h_{\pm}} \underset{b \rightarrow 0}{\longrightarrow} z^{\frac{1}{2} \pm \frac{m}{\epsilon_1}} \,,
\ee
where $h_{\pm}$ represents the conformal dimension of the two possible
fusion products of the degenerate operator with the mass insertion. In this limit, it is more appropriate to directly analyze
the $z \rightarrow 0$ limit of the differential equation (\ref{eq_in_w}) at
fixed $\epsilon_1$. This gives rise to the solutions ${\cal W} = \frac{1}{2}
(\epsilon_1 \pm 2 m) \log z$, indeed reproducing the expected behavior
near $z=0$.

\section{Conclusions}
\label{conclusions}
In this paper, we have studied consequences of the correspondence
\cite{Alday:2009aq} between two-dimensional conformal field theory and
${\cal N}=2$ supersymmetric gauge theories in four dimensions away
from the weak coupling limit on the gauge theory side.

Combining recursion relations satisfied by the toroidal conformal
blocks with modular results for $\cN=2^*$ and $\cN=4$ gauge theory
\cite{Huang:2011qx}, we have obtained new insights on both sides of
the correspondence. On the one hand, we have demonstrated that the
gauge theory results imply an infinite sequence of constraints,
non-perturbative in $q = e^{2 \pi i \tau}$, on the residua of the
conformal blocks. On the other, we have seen how the recursion
relation satisfied by the conformal block provides all genus results
for the topological string order by order in $q$. Furthermore, we have identified the scaling of the topological string partition function $F^{(n,g)}$ 
in the scalar vacuum expectation value $a$ in the field theory limit, $F^{(n,g)} \sim a^{2-2(g+n)}$,
as the coefficients, order by order in $q$, of a geometric series in $\frac{1}{a^2}$. The resulting
poles are exactly cancelled by zeros in the contribution to the
partition function stemming from maps that do not wrap the base of
the engineering geometry.

We have further demonstrated that the holomorphic anomaly results for
$\cN=2^*$ in the semi-classical limit can be reproduced by a null
vector decoupling equation of conformal field theory. The modularity
properties, while manifest in the former approach, arise highly
non-trivially in the solutions to this equation. The action of the
modular group maps the intermediate conformal dimension off the real
axis. We are thus led to consider one-point conformal blocks
parameterized by complex intermediate conformal dimension (the image
of the complex adjoint vacuum expectation value $a$ on the gauge
theory side). These form a larger class of conformal blocks than the
class that underlies unitary (Liouville) conformal field
theory. Unlike the traditional case, they exhibit modularity before
being assembled into a physical one-point function correlator. This
behavior is complementary to the intricate modular behavior of the
subset of blocks that close into each other under the S-move in
unitary Liouville theory \cite{Teschner:2003at,Hadasz:2009sw}. A detailed
perturbative analysis of the null vector decoupling equation on the
one-point toroidal conformal block allowed us to describe the
solutions in terms of a recursion procedure, providing an explicit
answer to all orders in $q$ in the expansion parameters $\epsilon_1$
and $m/a$, with $m$ determining the conformal dimension of the
insertion and $a$ the propagating momentum.

Our analysis allows for many generalizations to other ${\cal N}=2$
theories. A prime candidate is ${\cal N}=2$ $SU(2)$ gauge theory with
$N_f =4 $ fundamental flavors to which we hope to return in the near
future. 

We  believe that our analysis is one more indication
that complexifying parameters in conformal field theories with continuous
spectra is fruitful, as for instance indicated by the analysis of
the structure of Verma modules \cite{Feigin:1982tg}, the appearance of
interesting (analytically continued) operators for which there is no
corresponding state in the theory \cite{Seiberg:1990eb}, the analysis
of analytically continued operator product expansions and correlation
functions
\cite{Teschner:1999ug}\cite{Harlow:2011ny}
and
 the role of discrete states in the modularity
of theories with continuous spectrum \cite{Hanany:2002ev}\cite{Troost:2010ud}.

Further avenues for exploration are to explain the intriguing
analytical structure found for the topological string amplitudes directly in
that framework.
It would also be interesting to analyze the
link between the modular properties and the wave-function
interpretation of the conformal block with insertions in the analytic
continuation of Chern-Simons theory further. Finally, the occurrence of
quasi-modular forms in our analysis begs the questions whether their
almost holomorphic brethren also play a role in this context, and
whether the holomorphic anomaly equations can be derived purely within
conformal field theory.

\section*{Acknowledgements}
We would like to thank Jaume Gomis, Alessandro Tanzini 
and
Don Zagier 
for interesting discussions and correspondence.
Our work is supported in part by the grant ANR-09-BLAN-0157-02.

\appendix

\section{Properties of modular and elliptic functions}
In this appendix, we collect useful identities and modular and other properties of
modular forms and elliptic functions.
\subsection{Identities§}
The following identities hold:
\begin{eqnarray}
\wp(z) 
&=&  (\theta_1'/\theta_1)^2 - \theta_1''/\theta_1  - 2 \eta_1 \,,
\nonumber \\
\eta_1 &=& -2 \pi i \, \partial_\tau \log \eta(\tau) = \frac{\pi^2}{6} E_2(\tau) \,,
\nonumber \\
\partial_z^2 \theta_1  &=&  4 \pi i \partial_\tau \theta_1 \,.
\end{eqnarray}
The $\eta$-function and $\theta$-function have the following modular
and elliptic properties:
\be
\eta(\tau+1) = e^{\frac{i \pi }{12}} \eta(\tau) \,, \quad \eta(-\frac{1}{\tau}) = \sqrt{-i \tau} \,\eta(\tau)  \,,
\ee
\be \label{theta1_modular}
\theta_1(z|\tau+1 ) = e^{\frac{\pi i}{4}} \theta_1(z|\tau)   \,, \quad\theta_1(\frac{z}{\tau},-\frac{1}{\tau}) = -i \sqrt{-i \tau} e^{\frac{\pi i z^2}{\tau}}\theta_1(z|\tau) \,,
\ee
\be
\theta_1(z+1|\tau) = - \theta_1(z|\tau) \,, \quad \theta_1(z+\tau|\tau) = - e^{-\pi i \tau} e^{-2 \pi i z} \theta_1(z|\tau) \,.
\ee
The modular behavior of the second Eisenstein series is
\begin{eqnarray}
E_2 ( \frac{a \tau+b}{c \tau +d} ) &=&
(c \tau +d)^2 E_2 (\tau) + c (c \tau +d) \frac{6}{\pi i} \,.
\end{eqnarray}
The derivatives of the  Eisenstein series are
\begin{eqnarray}
q \partial_q E_2 = \frac{E_2^2 - E_4}{12}  \,,
\qquad
q \partial_q E_4 = \frac{E_2 E_4-E_6}{3} \,,
\qquad
q \partial_q E_6 = \frac{E_2 E_6 - E_4^2}{2} \,.
\end{eqnarray}

\subsection{The monodromies of powers of the Weierstrass function}
\label{monowp}
In this subsection, we determine the monodromies of powers of the 
Weierstrass function $\wp$
along a cycle $\gamma$ of an elliptic curve \cite{Halphen,Grosset}.
The elliptic curve is given by the equation $Y^2= 4 X^3 - g_2 X -g_3$.
The periods along $\gamma$ of a basis of differentials of the second kind will
be denoted by
\begin{eqnarray}
\omega &=& \int_\gamma dz = \frac{1}{2} \oint \frac{dX}{Y} \,,
\nonumber \\
\xi &=& - \frac{1}{2} \oint \wp(z) dz = - \frac{1}{2} \oint \frac{X dX}{Y}\,.
\end{eqnarray}
We can then write the monodromies of powers of the Weierstrass function $\wp$
along the cycle $\gamma$,
\begin{eqnarray}
K_n &=& \oint_\gamma \wp^n(z) dz \,,
\end{eqnarray}
as a linear combination
\begin{eqnarray}
K_n &=& A_n(g_2,g_3) \omega - B_n(g_2,g_3) \xi,
\end{eqnarray}
where $A_n$ and $B_n$ are polynomials of degree $2n$ and $2n-2$
respectively, where $g_2$ has degree $4$ and $g_3$ has degree $6$, and
the coefficients of the polynomials are positive and rational.
Moreover, they can be determined from the 
initialization
\begin{eqnarray}
& &  K_0 = 2 \omega \,,\qquad \qquad
K_1 = - 2 \xi \,,\qquad \qquad 
K_2 = \frac{1}{6} g_2 \omega \,,
\end{eqnarray}
and the recursion relation
\begin{eqnarray}
(8n-4) K_n &=& (2n-3) \, g_2 \, K_{n-2} + (2n-4) \, g_3 \, K_{n-3} \,.
\end{eqnarray}  
The first few integrals of powers of the Weierstrass function are
\begin{eqnarray}
K_3 &=& \frac{1}{10} ( 2 g_3 \omega - 3 g_2 \xi) \,,
\nonumber \\
K_4 &=& \frac{1}{168}(5 g_2^2 \omega -48 g_3 \xi)\,,
\nonumber \\
K_5 &=& \frac{1}{120} (8  g_2 g_3 \omega - 7 g_2^2 \xi) \,,
\nonumber \\
K_6 &=& \frac{1}{12320} ( (75 g_2^3+448 g_3^2) \omega - 1392 g_2 g_3 \xi) \,,
\nonumber \\
K_7 &=& \frac{1}{43680} ( 866 g_2^2 g_3 \omega - ( 539 g_2^3 + 2400 g_3^2) \xi)\,.
\end{eqnarray}
They can also be determined from the Halphen coefficients which
determine the powers of the Weierstrass function $\wp$ in terms of its
even derivatives. When $g_3=0$ or when $g_2=0$, the recursion
relation can be solved explicitly.

When we use these integrals in the bulk of the paper, we work on a torus parameterized by
a complex parameter $z$ where $z \equiv z+1 \equiv z + \tau$.  Moreover, we
restrict to determining the periods along the cycle $\alpha$ parameterized by $z \in [0,1]$.
Our initializations therefore read
\begin{eqnarray}
K_0 &=& \oint_\alpha dz = 1 = 2 \omega \,,
\nonumber \\
K_1 &=& \oint_\alpha \wp (z) dz = - 2 \eta_1 = - \frac{\pi^2}{3} E_2 = - 2 \xi \,,
\nonumber \\
K_2 &=&  \frac{\pi^4}{9} E_4\,.
\end{eqnarray}
We moreover have the relations
\begin{eqnarray}
g_2 &=& \frac{4 \pi^4}{3} E_4 \,,
\nonumber \\
g_3 &=& \frac{8 \pi^6}{27} E_6 \,. 
\nonumber \\
\end{eqnarray}
Under these circumstances, the first few period integrals 
of powers of the Weierstrass function are given by
\begin{eqnarray}
K_0 &=& 1 \,,
\nonumber \\
K_1 &=&  \frac{\pi^2}{3} (-E_2) \,,
\nonumber \\
K_2 &=& \frac{\pi^4}{9} E_4\,,
\nonumber \\
K_3 
 &=& \frac{\pi^6}{135} (-9 E_2 E_4 + 4 E_6)\,,
\nonumber \\
K_4 &=& \frac{\pi^8}{567}(15 E_4^2 -8 E_2 E_6) \,,
\nonumber \\
K_5 &=& \frac{\pi^{10}}{1215} ( -21 E_2 E_4^2 + 16 E_4 E_6) \,,
\nonumber \\
K_6 &=& \frac{\pi^{12}}{280665}
(2025 E_4^3 - 2088 E_2 E_4 E_6 + 448 E_6^2) \,,
\nonumber \\
K_7 &=& \frac{\pi^{14}}{995085}
( - 4851 E_2 E_4^3 - 800 E_2 E_6^2 + 5196 E_4^2 E_6) \,.
\end{eqnarray}

\bibliography{cft}
\bibliographystyle{utcaps}

\end{document}